\documentstyle[preprint,eqsecnum,aps,epsfig]{revtex}

\input epsf.sty
\newcommand{\be}{\begin{equation}}    
\newcommand{\ee}{\end{equation}}
\newcommand{\beq}{\begin{eqnarray}}
\newcommand{\eeq}{\end{eqnarray}}
\newcommand{\beqn}{\begin{eqnarray*}}
\newcommand{\eeqn}{\end{eqnarray*}}
\newcommand{\f}[2]{\frac{#1}{#2}}

\newcommand{\dps}{\displaystyle}        

\def\op{ \ $ }
\def\cl{$ \ }
\def\nn{\nonumber}

\def\ver{\vskip 12pt}
\def\ii{{\rm i}}   

\def\Y{Y_{l m}}

\def\Yt{\f{\partial Y_{l m}}{\partial\theta }}
\def\Ytt{\f{\partial^2 Y_{l m}}{\partial\theta^2 }}
\def\Yf{\f{\partial Y_{l m}}{\partial\phi }}

\def\EM{e^{-i\omega t}}

\def\N{N_{l m}}
\def\T{T_{l m}}
\def\V{V_{l m}}
\def\L{L_{l m}}
\def\X{X_{l m}}
\def\G{G_{l m}}
\def\hh{h^{0}_{l m}}
\def\hs{h^{1}_{l m}}
\def\pps{\Psi_{l m}}                                                                      
\def\IL{\relax{\rm I\kern-.18em L}}

\def\o{\omega}

\begin{document}


\draft

\title{Gravitational signals emitted
by a  point mass orbiting a neutron
star: a perturbative approach}

\author
{ L. Gualtieri$^1$, E. Berti$^{1}$, J.A. Pons$^{1}$, G. Miniutti$^{1}$
and V. Ferrari$^{1}$}
\address
{$^1$ Dipartimento di Fisica ``G.Marconi",
 Universit\` a di Roma ``La Sapienza"\\
and Sezione INFN  ROMA1, piazzale Aldo  Moro
2, I-00185 Roma, Italy}

\date{\today}

\maketitle

\begin{abstract}

We compute the energy spectra  of the gravitational 
signals emitted when a pointlike  mass  moves on a closed orbit 
around a non rotating neutron star, inducing a perturbation
of its gravitational field and its internal structure.
The Einstein equations and the hydrodynamical equations are perturbed and
numerically integrated in the frequency domain. 
The results are compared with the
energy spectra computed  by the quadrupole formalism which assumes
that both  masses are  pointlike, and
accounts only for  the radiation emitted 
because the orbital motion produces a time 
dependent quadrupole moment.
The results of our perturbative approach show that, in general, the quadrupole 
formalism overestimates the amount of emitted radiation, especially when 
the two masses are close.
However, if the pointlike mass  is allowed to move 
on an orbit  so tight that the keplerian orbital frequency 
resonates with  the frequency of the fundamental quasi-normal 
mode of the star ($2\omega_K=\omega_f$),
this mode  can be excited and the emitted 
radiation can be considerably larger than that 
computed by the quadrupole approach.
\end{abstract}
 

\narrowtext

\section{Introduction}

\ver\ver
The coalescence of binary systems composed of compact objects like black
holes or neutron stars is considered one of the most promising sources of
gravitational waves to be detected by ground-based interferometers.
For this reason it is important to collect as much information as possible
on the features of the gravitational signal  emitted in these
processes. This paper focuses on the phenomena which may occur
during the pre-merging phase of the coalescence, when the two stars are still
individual bodies in fast revolution around each other.
The problem  of computing the energy spectrum and the waveforms
of the gravitational waves emitted in this regime
can be attacked by using different approaches.
The easiest is  the quadrupole formalism, 
which assumes that the  two stars are pointlike masses, and
computes the emitted radiation  in terms of  the 
quadrupole moment of the system, whose time variation is due to
the orbital motion.
When the two stars get very close,
post newtonian corrections can be included to give a more
accurate description of the trajectories, and
to refine the orbital contribution of the emitted radiation.
These calculations can be complemented by the inclusion of radiation
reaction effects, which  account for the shrinking of the orbit
caused by the emission of gravitational waves. 

This approach clearly overlooks the fact that
the two stars are extended bodies with an internal structure and that,
when they are close,  the tidal interaction becomes strong and new
effects, unpredictable by the quadrupole + post-newtonian approaches, may arise.
An accurate description of these phases of the coalescence
requires the solution of the Einstein equations coupled with those of 
hydrodynamics in the non-linear regime, and many groups in the world 
are  working in this direction
\cite{nonlinear}.
These studies will certainly  yield new and interesting results,  but a complete picture 
is still far from reaching;
indeed, due to the complexity of this phenomenon 
the computational tools presently available allow  to follow the
evolution of the system for no more than a few orbits near coalescence.
Waiting for the results of fully non linear simulations,
it is interesting to explore
other techniques that, though approximated, allow to get some
insight into those phases of the coalescence where the quadrupole
+ post-newtonian approaches are inadequate. 
For instance, we can assume that one of the two stars is a
``true star", i.e. it is an extended body
whose equilibrium structure is described by 
a solution of the relativistic equations of  hydrostatic equilibrium,
and that only  the second star is a pointlike mass; its effect is to induce 
a perturbation on the gravitational field and on the
thermodynamical structure of the extended companion, which can be
evaluated by solving the equations of stellar perturbations in general
relativity.
In this way we can account for the tidal effects of the close
interaction on one of the two stars  and for  their
consequences on the gravitational emission, and get
a clue on the kind of phenomena that could arise near coalescence.
This approach has already been used in a previous paper
\cite{ferrarigualtieribor}  
where we computed  the energy spectrum of the
gravitational radiation emitted when a pointlike mass moves on an open
orbit around a compact star. 
In this paper we shall extend our investigation to the case of closed orbits,
either circular or eccentric,
and compute the energy spectra and the waveforms of the emitted radiation. 

The purpose of our study is   to  compare the quadrupole radiation emitted by
the system because of its orbital motion to  the signal
computed in the relativistic, perturbative approach.

The orbital emission  will be computed 
using a hybrid quadrupole approach, which assumes that the pointlike
mass moves on a geodesic of the spacetime generated by the star, but
radiates, according to the standard quadrupole formula, as if it were
in flat spacetime.

To evaluate the relativistic emission,
the equations of stellar perturbations we integrate in the
interior of the star  are those derived in ref. \cite{chandrafer1};
therefore we shall not describe them in detail, but only recall the
relevant formulae. 
In \cite{ferrarigualtieribor}, outside the star we integrated the 
Sasaki-Nakamura  equation \cite{nakamurasasaki}; here
we consider  closed orbits, i.e. a source with a compact support,
and it is more convenient
to integrate the Bardeen-Press-Teukolsky (BPT) equation
\cite{bardeenpress,teukolski} whose  source term has 
a much simpler form.

Since in this paper we do not consider the effects of radiation reaction,
we cannot describe the evolution of the orbit and of the waveform during
the inspiralling; thus the energy spectra we show  have to be
considered as representative of a certain number of orbital periods over
which the radiation reaction effects do not produce a significant
change in the orbital parameters of the pointlike mass
(adiabatic approximation).
These effects will be considered in a forthcoming paper.

Stellar perturbations excited by an orbiting particle
have also been considered by Kojima \cite{kojima},
who focused on the energy enhancement with respect
to the standard newtonian quadrupole formula, due to the
excitation of the fundamental mode  by a 
particle in circular orbit.
The excitation of the $w$-modes has been studied by
Ruoff, Laguna and Pullin
in a time domain approach \cite{ruofflagunapullin}.

The plan of the paper is the following.
In $\S$ II we write  the equations relevant to our problem,
in $\S$ III we discuss the source term of the BPT equation,
in $\S$ IV we outline the integration procedure, and discuss how to find the 
power emitted in gravitational waves and the waveforms. 
In $\S$ V we show how to compute the same quantities by the 
hybrid quadrupole approach.
Numerical results are discussed in $\S$ VI and conclusions are drawn 
in $\S$ VII.

\section{The perturbed equations}\label{model}

In order to describe the non-axisymmetric perturbations of a non rotating
star induced by an orbiting mass  we  choose, as in 
\cite{ferrarigualtieribor},  the following gauge
\footnote{The metric (\ref{pertmetric}) in \cite{ferrarigualtieribor}  
contained two misprints in the 
terms $d\theta d\phi$ and $dt d\phi$ which have now been corrected.}
\beq
\label{pertmetric}
ds^2 &=&
e^{2\nu(r)} dt^2 - e^{2\mu_2(r)} dr^2 -
r^2 d\theta^2-r^2\sin^2\theta d\phi^2\\
\nn
&+& \sum_{l m}~~\int^{+\infty}_{-\infty} d\omega~\EM
\Biggl\{
2 e^{2\nu}\N\Y dt^2-
2 e^{2\mu_2}\L\Y dr^2 - 2 r^2 H_{33} d\theta^2 - 
2 r^2\sin^2\theta H_{11} d\phi^2    \Biggr.\\
\nn
&-& \Biggl. 4r^2H_{13} d\theta d\phi 
+2\sin\theta\Yt  \left[ \hh dt d\phi
+ \hs dr d\phi\right]
- \f{2}{\sin\theta}\Yf\left[ \hh dt d\theta
+ \hs dr d\theta\right]
\Biggr\}
\eeq
where  the perturbed metric functions $[\N,\L,\V,\T,\hh,\hs]$ 
are functions of $(\omega,r)$,
\op \Y(\theta,\phi)\cl are the scalar spherical harmonics, and
\beq
\label{hhh}
\nn
   H_{11}&=&\left[T_{l m}+
V_{l m} \left( \f{1}{\sin^2{\theta}}
     \f{\partial^{2}}{\partial \phi^{2}}
     +\cot{\theta}\f{\partial}{\partial \theta}
     \right) \right] \ Y_{l m} \\\nn
   H_{13}&=&V_{l m}\left[
\f{\partial^2}{\partial\phi\partial\theta}
-\f{\partial}{\partial\phi}\cot\theta\right] Y_{l m}\\\nn
   H_{33}&=&\left[T_{l m}+
V_{l m}\f{\partial ^{2}}
     {\partial \theta^{2}}\right]\ Y_{l m} \ .
\nn
\eeq
The functions \op \Bigl[\N, \L, \V,\T\Bigr], \cl are the radial part
of  the {\it polar} ({\it even}) metric components,
and \op \hh\cl and \op\hs\cl are  the {\it axial} ({\it odd})
part.

The unperturbed metric
functions \op\nu(r)\cl and \op \mu_2(r)\cl can be found by solving the 
equations of hydrostatic equilibrium (cfr.  \cite{ferrarigualtieribor}, 
Eqs. (2.2)-(2.4)) for an assigned equation of state.
As in \cite{ferrarigualtieribor}, we shall consider as a model 
a polytropic compact star,
\op p=K\epsilon^{n},\cl with \op K=100~km^2\cl and
\op n=2.\cl 
If the  central density  is chosen to be
\op \epsilon_c=3\cdot 10^{15}~g/cm^3,\cl
the radius and mass of the star are, respectively, \op R_s=8.86~km\cl and
\op M=1.266~M_\odot,\cl with a ratio \op R_s/M=4.7.\cl
Although this model is, to some extent, unrealistic, it is appropriate 
to understand the basic features of the problem we want to study.

Inside the star, it is convenient to replace 
the  perturbed metric functions $\V$ and $\T $
by two new functions
\beqn
X&=& nV\\
G&=&\nu_{,r}[\frac{n+1}{n}X-T]_{,r}+
\frac{1}{r^{2}}(e^{2\mu_{2}}-1)[n(N+T)+N]+\frac{\nu_{,r}}{r}(N+L)-
\\
&-&e^{2\mu_{2}}(\epsilon+p)N+
\frac{1}{2}\omega^{2}e^{2(\mu_{2}-\nu)}[L-T+\frac{2n+1}{n}X]\: .
\eeqn
where \op n=(l -1)(l+2)/2.\cl 
The  polar metric functions \op \Bigl[\N, \L, \X,\G\Bigr]\cl
can be found  by solving a set of linear coupled equations 
(cfr. \cite{chandrafer1}, Eqs. 72-75), from which the fluid perturbations 
have been eliminated
\beq
\label{poleq}
&&X_{,r,r}+\left(\frac{2}{r}+\nu_{,r}-\mu_{2,r}\right)X_{,r}+
\frac{n}{r^{2}}
e^{2\mu_{2}}(N+L)+\omega^{2}e^{2(\mu_{2}-\nu)}X=0,
\\\nn
&&(r^{2}G){,r}=n\nu_{,r}(N-L)+\frac{n}{r}(e^{2\mu_{2}}-1)(N+L)+r(\nu_{,r}
-\mu_{2,r})X_{,r}+\omega^{2}e^{2(\mu_{2}-\nu)}rX\; ,
\\\nn
&&-\nu_{,r}N_{,r}=-G+\nu_{,r}[X_{,r}+\nu_{,r}(N-L)]
+\frac{1}{r^{2}}(e^{2\mu_{2}}-1)(N-rX_{,r}-r^{2}G)\\
\nn
&&-e^{2\mu_{2}}(\epsilon+p)N
+\frac{1}{2}\omega^{2}e^{2(\mu_{2}-\nu)}
\left\{ N+L+\frac{r^{2}}{n}G+\frac{1}{n}[rX_{,r}+(2n+1)X]\right\},
\\\nn
&&-L_{,r}=(N+2X)_{,r}+\left(\frac{1}{r}-\nu_{,r}\right)(-N+3L+2X)+\\
\nn
&&+\left[\frac{2}{r}-(Q+1)\nu_{,r}\right]\left[ N-L+\frac{r^{2}}{n}G+
\frac{1}{n}(rX_{,r}+X)\right]\; ,\nonumber
\eeq
where \op Q=\left(\frac{\partial p}{\partial \epsilon}
\right)^{-1}_{s} .\cl 

The equations for the axial perturbation
can be combined into a single wave
equation by introducing a function \op Z^{ax}_{l m}(\omega, r)\cl 
related to the axial functions by the following equations
\be
h^0_{l m}=
-\frac{i}{\omega}\frac{d}{dr_*}\left( r Z^{ax}_{l m}\right) ,
\qquad\qquad
h^1_{l m}= -  e^{-2\nu}\left( r Z^{ax}_{l m}\right) ,
\ee
where \op r_*=\int_0^re^{-\nu+\mu_2}dr.\cl
The equation \op Z^{ax}_{l m}(\omega, r)\cl satisfies is
(cfr. \cite{chandrafer1}, Eqs. 148-149) 
\be
\label{axeq}
\frac{d^2Z^{ax}_{l m}}{dr_*^2}+\Biggl\{\omega^2-
\frac{e^{2\nu}}{r^3}\Biggl[l(l+1)r+r^3(\epsilon-p)-6m(r)\Biggr]\Biggr\}
Z^{ax}_{l m}=0,
\ee
which, outside the star, automatically reduces to the Regge-Wheeler equation
\cite{reggewheeler}.
The polar and axial equations (\ref{poleq}) and  (\ref{axeq}) 
are numerically integrated  from $r=0$, where we impose 
a regularity condition, up to the surface of the star $r=R_s$. There 
we compute the amplitudes of the  Zerilli function \cite{zerilli} 
\be
\label{zerfun}
Z^{pol}_{l m}(\omega, R_s)=\f{R_s}{nR_s+3M}\Bigl[
\frac{3M}{n} X_{l m}(\omega,R_s)-R_sL_{l m}(\omega,R_s)
\Bigr],
\ee                                      
and of the Regge-Wheeler function, \op Z^{ax}_{l m}(\omega, R_s),\cl 
and their first derivatives, which will be used to continue 
the solution outside the star.

To describe the perturbations of the Schwarzschild  spacetime prevailing 
outside the star we use  the BPT equation
\cite{bardeenpress,teukolski}
\be
\label{teukolsky}
\left\{\Delta^2\frac{d}{dr}\left[\frac{1}{\Delta}\frac{d}{dr}\right]+
\left[\frac{\left(r^4\omega^2+4i(r-M)r^2\o\right)}{\Delta}
-8i\o r-2n\right]\right\}\pps (\o, r)= -T_{l m}(\o, r),
\ee
where \op \Delta=r^2-2Mr,\cl and  the BPT function, \op \pps, \cl is related
to the perturbation of the Weyl scalar \op \delta\Psi_4\cl by 
\be
\label{psiquattro}
\pps (\omega, r)=\frac{1}{2\pi} 
\int
d\Omega~dt~e^{ i\omega t}~_{-2}S^\ast_{l m}(\theta,\phi)
\left[ r^4~\delta\Psi_4(t,r,\theta,\phi)\right],
\ee
where \op _{-2}S_{l m}(\theta,\phi)\cl is the
spin-weighted spherical harmonic
\beq
_{-2}S_{l m}(\theta,\phi)=
\frac{1}{2\sqrt{n(n+1)}}\Biggl[
\Ytt-\cot\theta\Yt+\frac{m^2}{\sin^2\theta}\Y\Biggr.\\
\nn
\Biggl. + \frac{2 m}{\sin\theta}\left(\Yt-\Y \cot\theta\right)
\Biggr].
\eeq
The advantage of using  the BPT equation is that
\op \delta\Psi_4\cl  is invariant under gauge transformations and infinitesimal 
tetrad rotations,  and  that the squares of its real and 
imaginary parts  are proportional
to  the energy flux  of the outgoing radiation at infinity in the two
polarizations. In addition, the source term
\op T_{l m}(\omega,r),\cl which will be discussed in detail in the next section,
has a much simpler form than  the source of the Sasaki-Nakamura equation 
which we used to study the gravitational emission in the case of open orbits in 
{\cite{ferrarigualtieribor}}.

At the surface of the star, where \op T_{l m}(\omega,r)=0,\cl
the relation between \op\pps,\cl 
\op Z^{pol}_{l m}\cl and \op Z^{ax}_{l m}\cl is
\beq
\label{relazione1}
\pps&=&\frac{r^3\sqrt{n\left(n+1\right)}}{4\omega}\left[
V^{ax} Z^{ax}_{l m}  +\left(W^{ax} +2i\omega\right)\Lambda_+ Z^{ax}_{l m}  \right]\\
\nn
&-&\frac{r^3\sqrt{n\left(n+1\right)}}{4}\left[
V^{pol} Z^{pol}_{l m}+\left(W^{pol}+2i\omega\right)\Lambda_+  Z^{pol}_{l m}
\right], \eeq
where
\op
\Lambda_+={d\over dr_*}+i\omega=
{\Delta\over r^2}{d\over dr}+i\omega,
\cl
and
\beqn
V^{ax}&=&{2\Delta\over r^5}\left[(n+1)r-3M\right]\\
W^{ax}&=&{2\over r^2}(r-3M)\\
V^{pol}&=&{2\Delta\over r^5(nr+3M)^2}
\left[n^2(n+1)r^3+3Mn^2r^2+9M^2nr+9M^3\right]\nn\\
W^{pol}&=&2{nr^2-3Mnr-3M^2\over r^2(nr+3M)}.
\eeqn
According to (\ref{relazione1}) we can write \op\pps\cl as a sum 
of an axial  and a polar  part, i.e.
\be
\label{pp1}
\pps=\pps^{~ax}+\pps^{~pol},
\ee
where
\beq
\label{pp2}
{\Psi}^{~ax}_{l m} &=&
\frac{r^3\sqrt{n\left(n+1\right)}}{4\omega}\left[
V^{ax} {Z}^{ax}_{l m}  +\left(W^{ax} +2i\omega\right)\Lambda_+ 
{Z}^{ax}_{l m}  \right]\\
\nn
{\Psi}^{~pol}_{l m} &=&  
-\frac{r^3\sqrt{n\left(n+1\right)}}{4}\left[
V^{pol} {Z}^{pol}_{l m}+
\left(W^{pol}+2i\omega\right)\Lambda_+  {Z}^{pol}_{l m}
\right]. 
\eeq
However, in Section 4 we will show  that
depending on the value of the harmonic indices $l$ and $m$,
only the polar or the axial part of \op\pps\cl have to be selected.

\section{The source term of the BPT equation}

We shall assume that the pointlike mass $m_0$
which excites the perturbations
of the star follows a geodesic of the unperturbed
spacetime on the equatorial plane, with energy \op E\cl and angular 
momentum $ L_z$, so that the geodesic equations are
\be
\label{geod}
\f{dt}{d\tau}=\f{E}{1-\f{2M}{r}},~~~~
\f{dr}{d\tau}\equiv \gamma=\pm\sqrt{ E^2-\left(1-\f{2M}{r}  \right)
\left( 1+\f{L_z^2}{r^2} \right)},
~~~~\f{d\varphi}{d\tau}=\f{L_z}{ r^2}.
\ee
For a closed orbit
the motion takes place between a periastron $r_P$ and an apoastron $r_A$,
roots of the equation $\gamma=0$ (a third root marks the onset of a 
plunging motion, that is not relevant for our study). 
We can define the semi-latus rectum $p$ and the eccentricity $e$ in terms 
of $r_P$ and $r_A$ through the relations:
\be
r_P=\f{pM}{1+e},~~~~r_A=\f{pM}{1-e}.
\ee
Both $p$ and $e$ are dimensionless, and they are, respectively, a measure of 
the size and of the degree of noncircularity of the orbit. 
Note that $0\leq e < 1$.
The orbit is periodic in the radial coordinate, and quasi-periodic in the
$\varphi$-coordinate, i.e.
\beq
\label{period}
&&r(t+\Delta t)=r(t)\\\nn
&&\varphi(t+\Delta t)=\varphi(t)+\Delta\varphi.
\eeq
The source term of the BPT equation
(\ref{teukolsky}) is  
\beq
\label{sourcedef}
T_{l m}(\o,r)&=&-2\sqrt{n(n+1)}r^4 T_{(n)(n) l m}(\o,r)
-2\sqrt{n}\Delta\Lambda_+{r^5\over\Delta} T_{(n)(\bar m) l m}(\o,r)\\
\nn
&-&{\Delta\over 2r}\Lambda_+{r^6\over\Delta}\Lambda_+r T_{(\bar m) (\bar m) l m}(\o,r)\,,
\eeq
and can be found as follows.
The stress-energy tensor of the orbiting mass  
\be
\label{stress}
T^{\mu\nu}=\sum_{k=-\infty }^{\infty}{
\f{4\pi m_0}{r^2\vert \gamma\vert }~
\f{dz^{\mu}}{d\tau}~
\f{dz^{\nu}}{d\tau}~
\delta\Bigl(t-t_k(r)\Bigr)
\delta^{(2)}\Bigl(\Omega-\Omega_k(r)\Bigr)},
\ee
where $t_k(r),~\Omega_k(r)$ are the  time
and angular position of the mass on the $k$th  semi-orbit,
is projected onto the Newman-Penrose tetrad,
$\left(l,n,m,\bar{m}\right)$, to find its tetrad components
\op
T_{(p)(q)}=T^{\mu\nu}e_{(p)\mu}e_{(q)\nu}.\cl
These  are subsequently expanded in the spin-weighted 
spherical harmonics $_{s}S_{lm}(\theta,\phi),$ 
and Fourier expanded, as follows
\be
\label{stress1}
{T}_{(p)(q) l m}\left(\omega,r\right)=\f{1}{2\pi}
\int{dtd\Omega e^{i\omega t}~_{s}S^*_{lm}\left(\Omega\right)
T_{(p)(q)}\left(t,r,\Omega \right)},
\ee
where $s=-2,-1,0~$ for $~T_{(\bar m)(\bar m)},  T_{(n)(\bar m)},
 T_{(n)(n)}~$ respectively.
As a result of this procedure we find
\be
{T}_{(p)(q) l m}\left(\omega,r\right)=\left(
\sum_{k=-\infty}^{+\infty}e^{ik\left[\omega\Delta t-m\Delta\varphi\right]}
\right){T}_{0(p)(q) l m}\left(\omega,r\right).
\label{stress2}
\ee
The  explicit expressions of \op {T}_{0(p)(q) l m}\cl are given in appendix A.
By making use of the relation
\be
\label{deltas}
\sum_{k=-\infty}^{+\infty}e^{ikX}=2\pi
\sum_{j=-\infty}^{+\infty}\delta\left(X-2\pi j\right),
\ee
it is easy to see that Eq. (\ref{stress2}) can be written as
\be
\label{stress3}
{T}_{(p)(q) l m}\left(\omega,r\right)=
{2\pi\over\Delta t} \sum_{j=-\infty}^{+\infty}
\delta\left(\omega-\o_{m j}\right){T}_{0 (p)
(q) l m}\left(\omega,r\right),
\ee
where 
\be
\label{freq}
\omega_{m j}={2\pi j+m\Delta\varphi\over\Delta t}
\equiv j\Omega_r+m\Omega_\varphi,
\ee
and the source of the BPT equation takes the form
\be
\label{source}
T_{l m}(\omega,r)=\sum_{j=-\infty}^{+\infty} \delta\left(\omega-\omega_{m j}\right)
T_{0~ l m}(\omega,r).
\ee
It should be noted that
$\Omega_r$ and $\Omega_\varphi$ are the two characteristic frequencies of
the problem. The frequency $\Omega_r=\f{2\pi}{\Delta t}$ is associated to the
periodicity of the radial motion, whereas 
$\Omega_\varphi=\f{\Delta\varphi}{\Delta t}$
is the angular velocity of an inertial observer  with respect to which
the $\varphi$-motion of the orbiting mass appears to be periodic.

\section{The solution of  the BPT equation}
The inhomogeneous BPT equation (\ref{teukolsky}) can be integrated by 
constructing a Green function which ensures that \op\delta\Psi_4\cl
matches  regularly with the interior solution
at the boundary of the star, and behaves as a pure outgoing 
wave at infinity. This problem has been solved by Detweiler in the case of
black holes \cite{det}; here we shortly describe the simple generalization 
of the method to the case of stars.
First we discuss some symmetry properties of  the functions involved in 
the problem we want to solve.
Under complex conjugation, spherical harmonics   behave as
\op 
Y^\ast_{l m}(\theta,\phi)=
(-1)^{m}Y_{l -m}(\theta,\phi);
\cl
consequently,
the perturbed metric functions in (\ref{pertmetric}) and the 
functions \op Z^{pol}_{l m}\cl and \op Z^{ax}_{l m},\cl
satisfy the following property
\op
F^\ast_{l m}(r,\omega)=(-1)^{m}F_{l -m}(r,-\omega).
\cl
From eqs. (\ref{pp2}) it follows that
\beq
\label{prop2}
\left({\Psi}^{~ax}_{l m}(\omega,r)\right)^\ast 
&=& (-1)^{m+1}~{\Psi}^{~ax}_{l -m}(-\omega,r)\\
\nn
\left({\Psi}^{~pol}_{l m}(\omega,r)\right)^\ast 
&=& (-1)^{m}~{\Psi}^{~pol}_{l m}(-\omega,r).  
\eeq
By inspection of the source term, we find that
\op\pps\cl must satisfy an additional relation
\be
\label{prop3}
\Psi^{~\ast}_{l m}(r,\omega)=(-1)^{l}\Psi_{l -m}(r,-\omega).
\ee
In order (\ref{prop2}) and (\ref{prop3}) be consistent, 
and looking at eqs. (\ref{pp1}) we see that the following selection rule must hold
\begin{itemize}
\item{}  if \op (l +m)\cl is even, 
\op \pps= {\Psi}^{~pol}_{lm} ,\cl
\item{}  if \op (l +m)\cl  is odd, 
\op \pps= {\Psi}^{~ax}_{lm} .\cl
\end{itemize}
Thus, depending on the value of the harmonic indices $l$ and $m$,
\op\pps\cl is either  polar or  axial.
As explained in Section II, we integrate the equations of stellar 
perturbations in the interior of the star (\ref{poleq}) and
(\ref{axeq}), and  construct the functions
\op Z^{ax}_{l m}(\omega, R_s)\cl and \op Z^{pol}_{l m}(\omega, R_s)\cl 
and their first derivatives at  $r=R_s$;  from them 
we compute \op {\Psi}^{~ax}_{lm}(\o,R_s)\cl and \op {\Psi}^{~pol}_{lm}(\o,R_s),\cl
as given in eqs. (\ref{pp2}), and their first derivatives, which are needed to integrate 
the BPT equation outside the star.
However, it should be noted that the regularity condition imposed at 
$r=0$ allows to determine \op Z^{ax}_{l m}\cl and \op Z^{pol}_{l m}\cl
only up to an unknown amplitude, \op\chi_{l m}(\o),\cl
to be determined by the matching conditions
at the boundary of the star. In what follows, we shall indicate as
\op\bar{\Psi}^{~ax}_{lm}(\o,R_s)\cl and \op\bar{\Psi}^{~pol}_{lm}(\o,R_s)\cl
the values of the axial and polar part of the wavefunction \op\pps\cl 
as computed by numerical integration of the interior equations.
The problem we want to solve therefore is
\beq
\label{system}
&&\IL_{BPT}\pps(\omega,r)=-T_{l m}(\omega,r)\\
\nn
&&\pps(r\rightarrow\infty)=r^3e^{\ii\o r_*}A_{l m}(\o)\\
\nn
&&\pps(\o,R_s)=\chi_{l m}(\o){\bar\pps}(\o, R_s) \\
\nn
&&\Psi^{'}_{l m}(\o,R_s)=\chi_{l m}(\o){\bar\Psi}^{'}_{l m}(\o, R_s),
\nn
\eeq
where $\IL_{BPT}$ is the differential operator on the left hand side
of the BPT equation. 
If   \op (l +m)\cl is even, 
\op {\bar\pps}(\o, R_s) =  {\bar\Psi}^{~pol}_{lm}(\o,R_s),\cl 
whereas if  \op (l +m)\cl is odd
\op {\bar\pps}(\o, R_s)={\bar\Psi}^{~ax}_{lm}(\o,R_s).\cl 
\op A_{l m}(\o)\cl is the  unknown wave amplitude to be determined.
The general solution of eqs. (\ref{system}) is
\be
\label{solution}
\pps(\o,r)=-{1\over W_{l m}}\left[
\Psi^{~0}_{l m}~
\int_{R_s}^r~ \frac{dr'}{\Delta^2}~\Psi^{~1}_{l m}~ T_{l m}
+
\Psi^{~1}_{l m}
\int_{r}^{\infty}~ \frac{dr'}{\Delta^2}~\Psi^{~0}_{l m}~ T_{l m}
\right],
\ee
where \op \Psi^{~0}_{l m}\cl and \op \Psi^{~1}_{l m}\cl
are two independent solutions of the homogeneous BPT equation
defined as
\be
\label{Ro}
\cases{\dps{
\IL_{BPT}\Psi^{~0}_{l m}(\o,r)=0}&\cr
\dps{
\Psi^{~0}_{l m}(\o,r\rightarrow\infty) =r^3e^{\ii\o r_*}}&\cr
},\qquad\qquad
\cases{\dps{
\IL_{BPT}\Psi^{~1}_{l m}(\o,r)=0} &\cr
\dps{\Psi^{~1}_{l m}(\o,R_s)= \bar{\Psi}_{l m}(\o,R_s)}&\cr
\dps{\Psi^{~1~'}_{l m}(\o,R_s)= \bar{\Psi}^{~'}_{l m}(\o,R_s)}&\cr
},
\ee
and \op  W_{l m}(\o)\cl is the Wronskian
\be
W_{l m}(\o)= \frac{1}{\Delta}\left[
\Psi^{~1}_{l m} \Psi^{~0}_{l m~,r}-\Psi^{~0}_{l m} \Psi^{~1}_{l m~,r}
\right].
\ee
From eq. (\ref{solution}) it is easy to see that the amplitude of the wave at infinity
is
\be
\label{ampli0}
A_{l m}(\o)=
-{1\over W_{l m}(\o)}~
\int_{R_s}^{\infty}~ \frac{dr'}{\Delta^2}~\Psi^{~1}_{l m}(\o,r')~ T_{l m}(\o,r').
\ee
By the use of eq. (\ref{source}) this expression becomes
\be
\label{ampli1}
A_{l m}(\o)=
-{1\over W_{l m}(\o)}~
{2\pi\over\Delta t} \sum_{j=-\infty}^{+\infty}
\delta\left(\omega-\omega_{m j}\right)
\int_{R_s}^{\infty}~ \frac{dr'}{\Delta^2}~\Psi^{~1}_{l m}(\o,r')~ 
T_{0~l m}(\o,r').
\ee
Some further details  related to the evaluation of the amplitude
(\ref{ampli1}) are given in Appendix A.

We shall now compute the time-averaged energy-flux
\be
\label{enflux}
\left<dE\over dt\right>=\lim_{T\rightarrow\infty}{E\over T}
=\lim_{T\rightarrow\infty}{1\over T}\sum_{lm}\int d\omega
\left(dE\over d\omega\right)_{lm},
\ee
where the  energy spectrum, $\frac{dE}{d\omega}$, can be  expressed 
in terms of the wave amplitude at infinity \op A_{lm}(\omega)\cl
as
\be
\left(\f{dE}{d\omega}\right)_{lm}=
{1\over 2\omega^2}\vert A_{lm}(\omega)\vert^2.
\ee
Since the wave amplitude can be written as (cfr. Eq. \ref{ampli1})
\be
\label{ampli}
A_{l m}(\omega)
=\sum_{j=-\infty}^\infty \hat A_{l m}(\omega)\delta(\omega-\omega_{m j})
\ee
we have
\beq
\label{amplispec}
\left<dE\over dt\right>
&=&\lim_{T\rightarrow\infty}{1\over T}~
\sum_{lm} 
\sum_{j=-\infty}^{+\infty}
\int d\omega~
{\rm ''}\delta(\omega-\omega_{m j})^2{\rm ''}~{1 \over 2\omega^2}
\vert \hat A_{lm}(\omega)\vert^2\\\nn
&=&
\sum_{lm} \sum_{j=-\infty}^{+\infty}~
\f{1}{4\pi \omega_{m j}^2}~
\vert \hat A_{lm}(\omega_{m j})\vert^2
\equiv \sum_{lm}
\sum_{j=-\infty}^{+\infty}
\dot{E}^R_{lmj},
\eeq
where ${\rm ''}\delta^2{\rm ''}$  is the regularized squared $\delta$-function,
such that
\be
\lim_{T\rightarrow\infty}{2\pi\over T}
{\rm ''}\delta(\omega-\omega_{m j})^2{\rm ''}
=\delta(\omega-\omega_{m j}),
\ee
and we have defined the time-averaged power spectrum
\be\label{relflux}
\dot{E}^R_{lmj}\equiv 
{1\over 4\pi \omega_{m j}^2}
\vert \hat A_{lm}(\omega_{m j})\vert^2.
\ee
In conclusion, the gravitational emission is characterized 
by a series of spectral lines at frequencies \op \omega_{m j}.\cl 
From  the symmetry properties 
\be
\label{symmOmega}
\omega_{-j-m}=-\omega_{m j}
\ee
and
\be\label{symmA}
\hat A^\ast_{lm}(\omega_{m j})=(-1)^l~
\hat A_{l-m}(\omega_{-j-m}), \qquad j>0
\ee
it follows that 
\be\label{energyparity}
\dot{E}^R_{l -m -j}=\dot{E}^R_{l m j}.
\ee
Thus, once we know the power spectrum $\dot{E}^R_{l m j}$ 
as a function of the frequencies 
$\omega_{m j}$, 
for an assigned value of  $l$ and for positive $m,$ 
the spectrum  for negative $m$ is obtained by  Eq. (\ref{energyparity}).

Since  $\delta\Psi_4(t,r,\theta,\phi)$ and the gravitational
wave amplitude in the  radiation gauge are related by
\be
\delta\Psi_4(t,r,\theta,\phi)
=-{1\over 2}\left[
\ddot h^{TT}_+(t,r,\theta,\phi)+\ii\ddot h^{TT}_\times(t,r,\theta,\phi)
\right]
\ee
using  Eq. (\ref{psiquattro}) we find
\beq
\label{onda1}
\left[r h^{TT}_{+~ l m}\left(t,r,\theta,\phi\right)\right]_{r\rightarrow\infty}
&=&
~_{-2}S_{lm}\left(\theta,0\right) ~
{\rm Re}~
e^{im\phi}
\sum_{j=-\infty}^\infty~
\frac{2}{\omega^2_{m j} }~
\hat A_{lm}(\omega_{m j})e^{-i\omega_{m j}(t-r_*)}\\
\nn
\left[r h^{TT}_{\times~l m}
\left(t,r,\theta,\phi\right)\right]_{r\rightarrow\infty}&=&
~_{-2}S_{lm}\left(\theta,0\right) 
{\rm Im}~
e^{im\phi}
\sum_{j=-\infty}^\infty~
{2\over \omega_{m j}^2 }~
\hat A_{lm}(\omega_{m j})e^{-i\omega_{m j}(t-r_*)}\nn
\eeq
where we have separated the $\phi-$dependence of the spin weighted 
spherical harmonics,
\op _{-2}S_{l m}(\theta,\phi)=_{-2}S_{l m}(\theta,0)~e^{im\phi}.\cl

\section{The  quadrupole emission }
We shall now compute the gravitational radiation emitted by the
mass $m_0$  because of its accelerated orbital motion around the star.
The energy flux  is computed by using a semi-relativistic approximation, 
which assumes that  $m_0$  moves along a geodesic of the 
curved spacetime, but radiates as if it were in flat spacetime. 
Using the quadrupole formula, it is easy to show that
the TT-components of the gravitational wave 
emitted by the particle are  \cite{kostas}
\beq \label{htp}
r h^{TT}_{\theta\theta}(t,r,\theta,\phi) &=&
 \left[\left(\ddot{Q}_{xx}-\ddot{Q}_{yy} \right)\cos^2\phi +
\ddot{Q}_{xy} \sin2\phi
 \right] \left( 1+\cos^2\theta\right) \nn \\
&+&\left[\ddot{Q}_{yy}-\ddot{Q}_{zz}  \right]\cos^2\theta +
\ddot{Q}_{zz}- \ddot{Q}_{xx}\\\nn
r h^{TT}_{\theta\phi}(t,r,\theta,\phi)
&=&-\cos\theta \left[ \left(\ddot{Q}_{xx}-\ddot{Q}_{yy}\right)
\sin2\phi+ 2\ddot{Q}_{xy}\left(1-2\cos^2\phi\right)
\right]
\eeq
where \op Q_{kl}\cl denotes
the components of the reduced quadrupole moment,
\op Q_{kl}=m_0\left(X^{k}X^{l}-\frac{1}{3}\delta^{k}_{l}\left|{\mathbf{X}}
\right|^{2}\right), \cl and
\op\theta,\phi\cl are the polar angles.
The two-dimensional vector $ {\mathbf{X}}$ is the position of the
particle along its trajectory in the equatorial plane 
$ {\mathbf{X}}= \left( r(t)\cos\varphi(t), r(t)\sin\varphi(t) \right)$, and 
\op r(t)\cl and \op \varphi(t)\cl are given by the geodesic
equations (\ref{geod}). 
The expressions of the second time-derivative of the
components of \op Q_{kl}\cl in terms of \op r(t)\cl and \op \varphi(t)\cl
are
\beq \label{qxx}
\ddot{Q}_{xx} &=&  
m_0\left(\alpha\cos2\varphi-\beta\sin2\varphi+\delta/2\right),\nn\\
\ddot{Q}_{yy} &=&m_0\left(-\alpha\cos2\varphi+\beta\sin2\varphi+\delta/2\right),\\
\ddot{Q}_{xy} &=&m_0\left(\alpha\sin2\varphi+\beta\cos2\varphi\right),
\quad
\ddot{Q}_{zz} =-m_0\delta,  
\nn
\eeq
where 
\beq
\label{alphabeta}
&&\alpha = \dot{r}^2+r\ddot{r}-2r^2\dot{\varphi}^2,\\\nn
&&\beta = 4r\dot{r}\dot{\varphi}+r^2\ddot{\varphi},\quad
\delta= 2 \left(\dot{r}^2+r\ddot{r}\right)/3.
\eeq
In  Appendix B we explicitely 
compute the Fourier transform of the metric components 
(\ref{htp}), which will be used to evaluate the energy flux, 
and we show that they can be written as
\beq
\label{htp1}
h^{TT}_{\theta \theta}(\omega, r, \theta, \phi)&=&
\sum_{m=-2,0,2}~
\sum_{j=-\infty}^{\infty}~
\delta(\omega-\omega_{m j})~H_{\theta \theta}(\omega_{m j},\theta,\phi),
\\\nn
h^{TT}_{\theta \phi}(\omega, r, \theta, \phi)&=&
- \sum_{m=-2,2}~ 
\sum_{j=-\infty}^{\infty}
\delta(\omega-\omega_{m j})~H_{\theta \phi}(\omega_{m j},\theta,\phi),
\eeq
where \op \omega_{m j}\cl are defined in eq. (\ref{freq}), and 
\op H_{\theta \theta}, H_{\theta \phi}\cl are  given in Appendix B.
We shall now derive  the time-averaged quadrupole energy flux (\ref{enflux}).
Since
\be
\f{dE^{(Q)}}{dSdt}=\f{1}{16\pi} 
\left\{
\left|\dot{h}^{TT}_{\theta \theta}(t,r,\theta,\phi)\right|^2+
\left|\dot{h}^{TT}_{\theta \phi}(t,r,\theta,\phi)\right|^2
\right\}
\ee
 it follows that
\be
\left\langle \f{dE^{(Q)}}{dt}\right\rangle
= \f{r^2}{16\pi}
\lim_{T\to \infty}\f{1}{T}\int_0^T dt \int d \Omega
\left\{
\left|\dot{h}^{TT}_{\theta \theta}\right|^2+
\left|\dot{h}^{TT}_{\theta \phi}\right|^2
\right\};
\ee
using  Parseval's theorem this becomes
\beq
\label{quadflux}
\left\langle \f{dE^{(Q)}}{dt}\right\rangle
&=&
\frac{r^2}{16\pi}
\sum_{j=-\infty}^{\infty}
\left[
\sum_{m=-2,0,2}
\int{d\Omega~
\omega^2_{m j}~\vert H_{\theta \theta}(\omega_{m j},\theta,\phi)\vert^2+
\sum_{m=-2,2}
}
\int{d\Omega~
\omega^2_{m j}~\vert H_{\theta \phi}(\omega_{m j},\theta,\phi)\vert^2
}
\right]
\nn
\\
&=& \sum_{j=-\infty}^{\infty}~\sum_{m=-2,0,2} ~\dot{E}^Q_{m j},
\eeq
where 
\beq
&\hbox{for $m=-2,2$}\qquad\qquad
&\dot{E}^Q_{m j}=
\frac{r^2}{16\pi} \int{d\Omega~\left[
\omega^2_{m j}~\vert H_{\theta \theta}\vert^2+
\omega^2_{m j}~\vert H_{\theta \phi}\vert^2
\right]},\\
\nn
&\hbox{for $m=0$}\qquad\qquad
&\dot{E}^Q_{m j}=
\frac{r^2}{16\pi} \int{d\Omega~\left[
\omega^2_{m j}~\vert H_{\theta \theta}\vert^2
\right]}.
\eeq

\section{Numerical results}
The equations of stellar perturbations (\ref{poleq}), (\ref{axeq}) and (\ref{teukolsky})
have been numerically integrated  for a set of
bounded orbits identified by selected values of the orbital parameters
$(E,L_z)$, or, equivalently,  $(e,r_P)$. 
In computing the energy flux, 
we have seen that the energy is emitted at a discrete, infinite set of frequencies 
\op\omega_{m j},\cl with \op-\infty < j < +\infty,\cl defined in eq. (\ref{freq}). 
The output of our perturbative calculations are the amplitudes of the 
spectral lines \op \dot{E}^R_{l m j}\cl
(\ref{relflux}) and the corresponding waveforms (\ref{onda1}).

The energy computed by the
hybrid quadrupole approach is also emitted at the same  discrete 
frequencies \op\omega_{m j};\cl 
however, whereas the quadrupole emission is resticted to 
$l=2$, and $m=(-2,0,2)$ for \op h^{TT}_{\theta \theta}\cl 
and  $m=(-2,2)$ for \op h^{TT}_{\theta \phi}\cl  (cfr. Eqs.  \ref{htp1}),
for the relativistic calculations $l \ge 2$ 
and $ -l < m < l$ for both polarizations.
Thus to compare the outcome of the two approaches we have to
confront the quadrupole spectral lines \op \dot E^{Q}_{m j}\cl
with the $l=2$  relativistic lines \op \dot{E}^R_{l m j}.\cl

\begin{figure}
\centerline{\mbox{
\epsfxsize=10cm \epsfbox{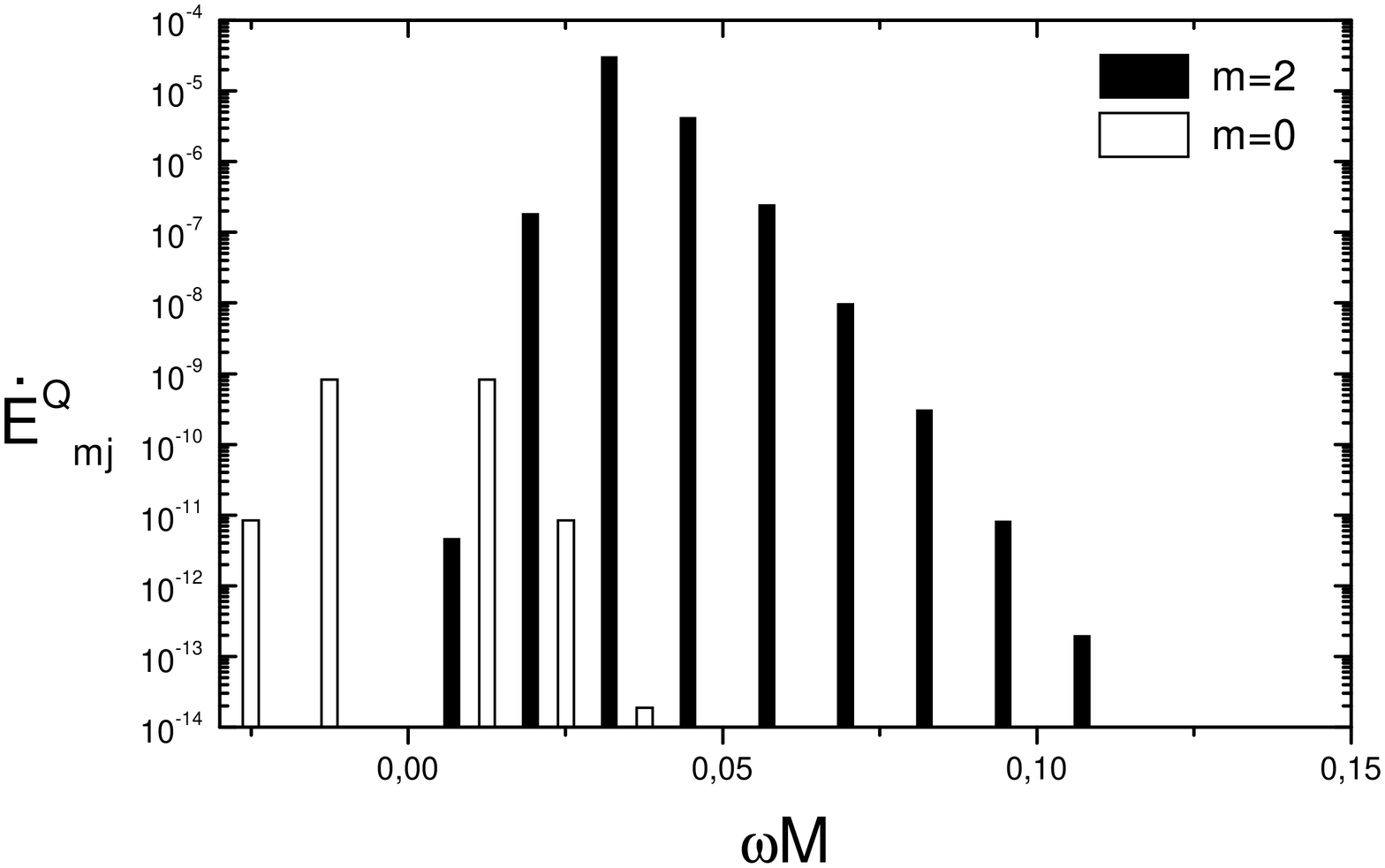}
\epsfxsize=10cm \epsfbox{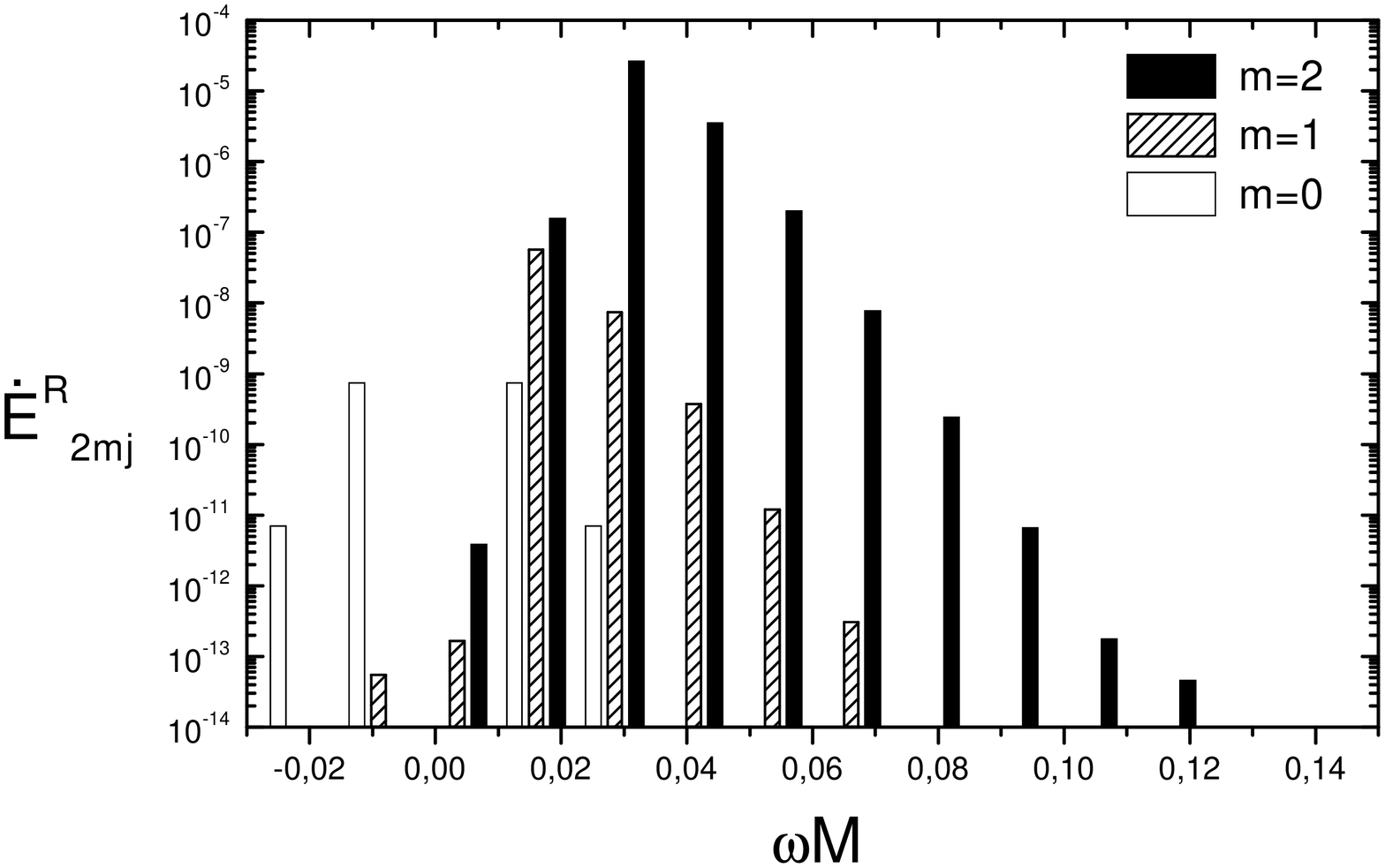}
}}
\vskip 8pt
\centerline{\mbox{      
\epsfxsize=10cm \epsfbox{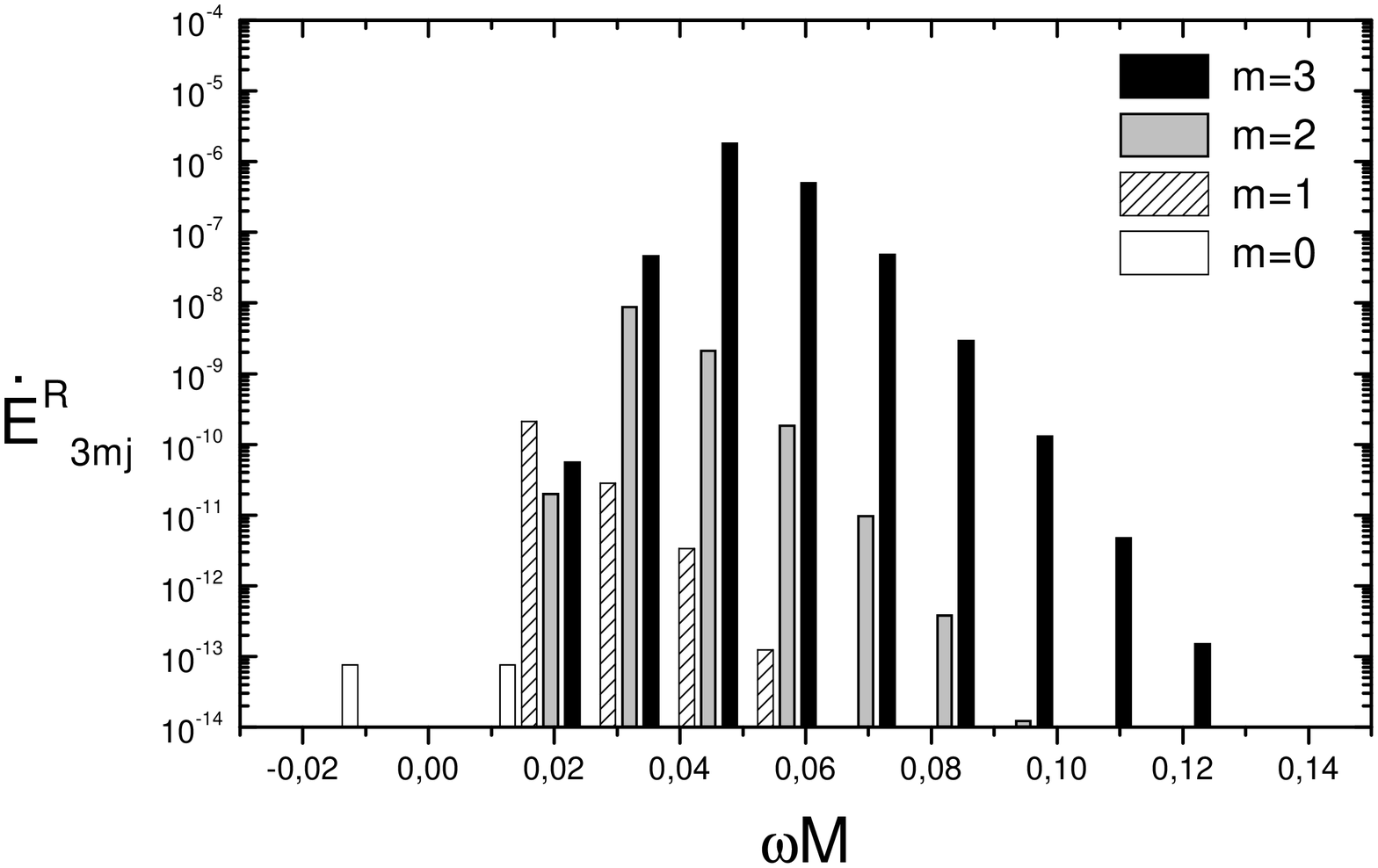}
\epsfxsize=10cm \epsfbox{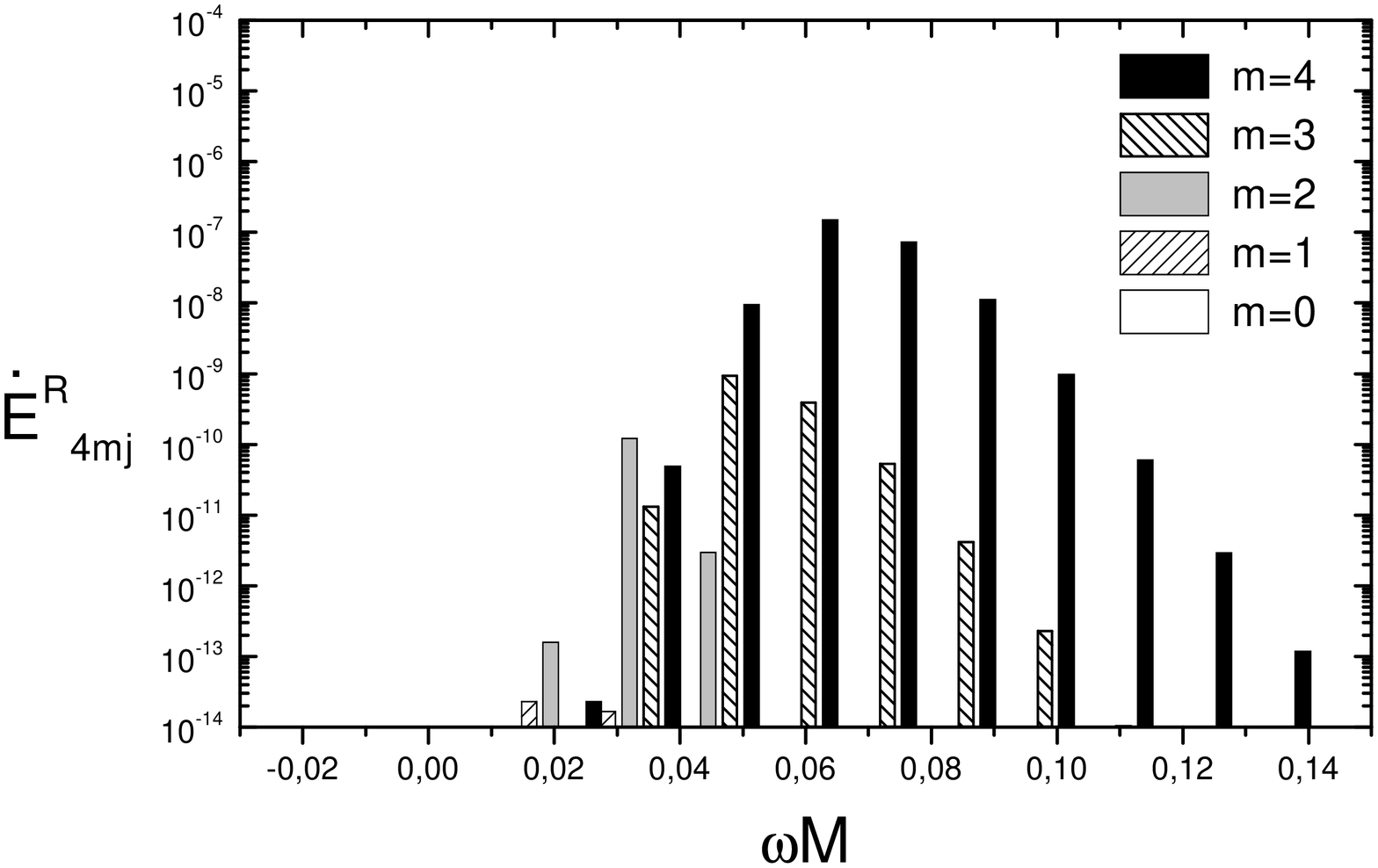}
}}
\vskip 10pt
\caption{
The gravitational emission associated to an eccentric orbit
with  $e=0.1$ and $r_P=3R_s,$ is illustrated  by plotting the amplitude of 
the spectral lines versus the dimensionless frequency \op M\omega_{m j}. \cl
The radiation computed by the hybrid quadrupole approach \op \dot E^Q_{m j}\cl
(upper panel, left) has to be compared 
with that computed by the relativistic approach,
\op \dot E^R_{l m j },\cl  for \op l=2\cl (upper panel, right). In the lower 
panel we plot the relativistic   $l=3$ and $l=4$ contributions.
}
\label{mdependence}
\end{figure}

In Figure \ref{mdependence} we show, as an example, the energy output 
for an orbit with periastron $r_P=3R_s$ and eccentricity $e=0.1$ 
computed by the quadrupole approach,
and  the relativistic results
for $l=2,~3$ and $4,$  for the same orbit.
The spectral lines are plotted for  the discrete
values of the dimensionless frequency $M\omega_{m j} $, 
for assigned values of positive $m$.  
We do not plot the lines corresponding to  negative $m$   because 
they can be obtained through a reflection across the zero frequency axis
of the positive ones, by virtue of the symmetry property (\ref{energyparity}).
A comparison of the quadrupole emission (upper panel, left) with the $l=2$ 
relativistic emission (upper panel, right),
shows that for $m=0$ and $m=2$ the two spectra are
qualitatively similar.
As expected, the three plots
which refer to the perturbative results
show that most of the energy is emitted in the $l=2$ multipole,
and that for each $l$, the $l=m$ component is always 
larger than the others.

\begin{figure}[htbp]
\begin{center}
\leavevmode
\epsfxsize=16cm \epsfbox{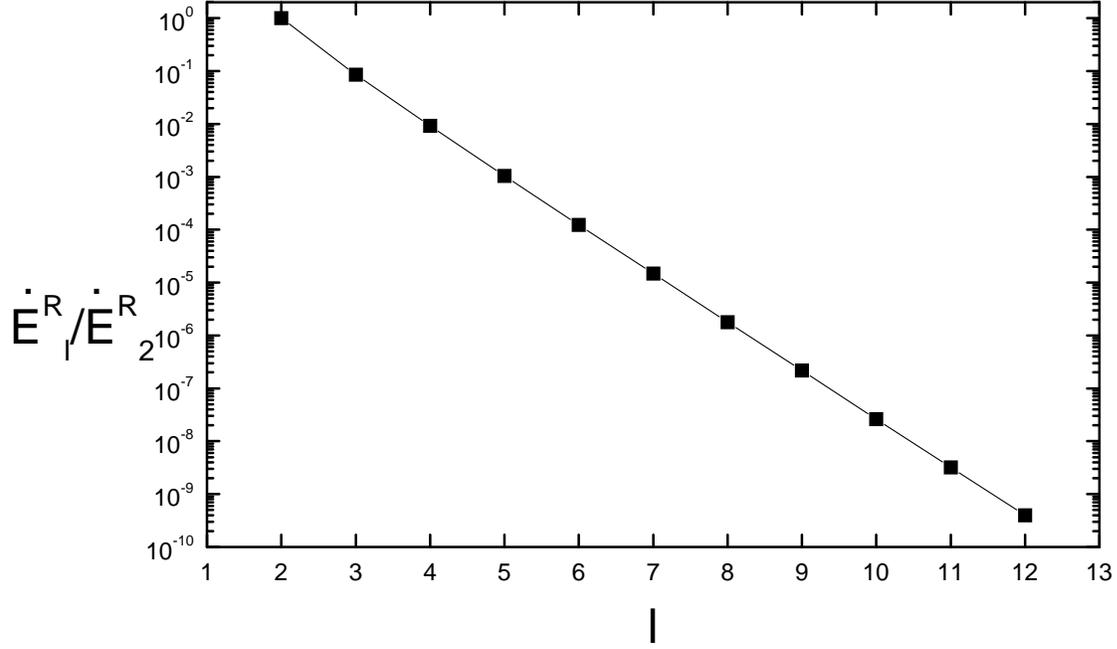}
\end{center}
\caption{
The ratio of the total power emitted  in the $l$-th multipole, \op\dot E^R_l,\cl
to the total power emitted in $l=2$, \op \dot E^R_2,\cl
is plotted as a function of $l$,
for an orbit with $e=0.1$ and $p=15.64$ ($r_P=3~R_s$).
It clearly shows  a power law behaviour  
\op\dot E^R_l\sim p^{l-2}.\cl 
}\label{multipoles}
\end{figure}

It is known that
for particles in circular orbit around black holes the total power 
emitted in each multipole 
\be\label{totpow}
\dot E^R_l = \sum_{j=-\infty}^\infty\sum_{m=-2}^2 \dot E^R_{lmj}
\ee
scales with the multipole order as 
\be\label{scalelaw}
\dot E^R_l\sim p^{2-l},
\ee
where $p$ is the orbital semi-latus rectum \cite{Poisson}.
This power-law scaling was found analytically.
Subsequently, 
Cutler, Kennefick and Poisson \cite{CutlerKennefickPoisson} numerically
integrated the BPT equation for a Schwarzschild
black hole with a point particle moving on bounded orbits.
They showed that the same result  holds, at least in order of magnitude,  
also for particles in eccentric orbits.
We find that
a similar power law exists also for stars, both for circular and eccentric
orbits, as indicated in Figure \ref{multipoles} where, as an 
example, we plot the ratio $\dot E^R_l/\dot E^R_2$ as a function of $l$,
for an orbit with $e=0.1$ and $p=15.64$
($r_P=3~R_s$).
In Table 1 we tabulate the ratio \op \dot E^R_l/\dot E^R_2\cl for different 
values of $l$ and for two circular orbits with \op r_0=3~R_s\cl and
\op r_0=10~R_s,\cl respectively.

\begin{figure}
\centerline{\mbox{
\epsfxsize=10cm \epsfbox{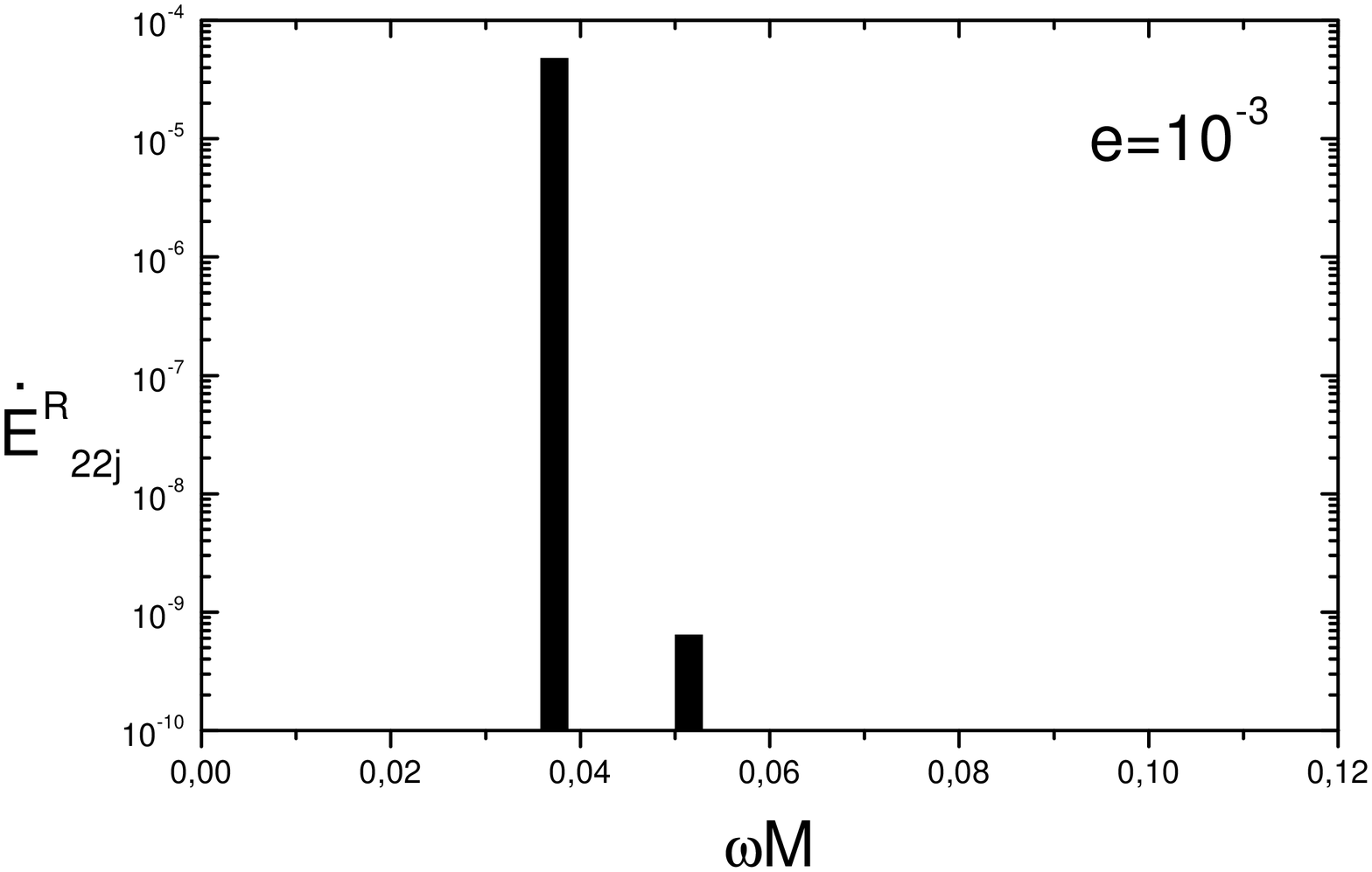}
\epsfxsize=10cm \epsfbox{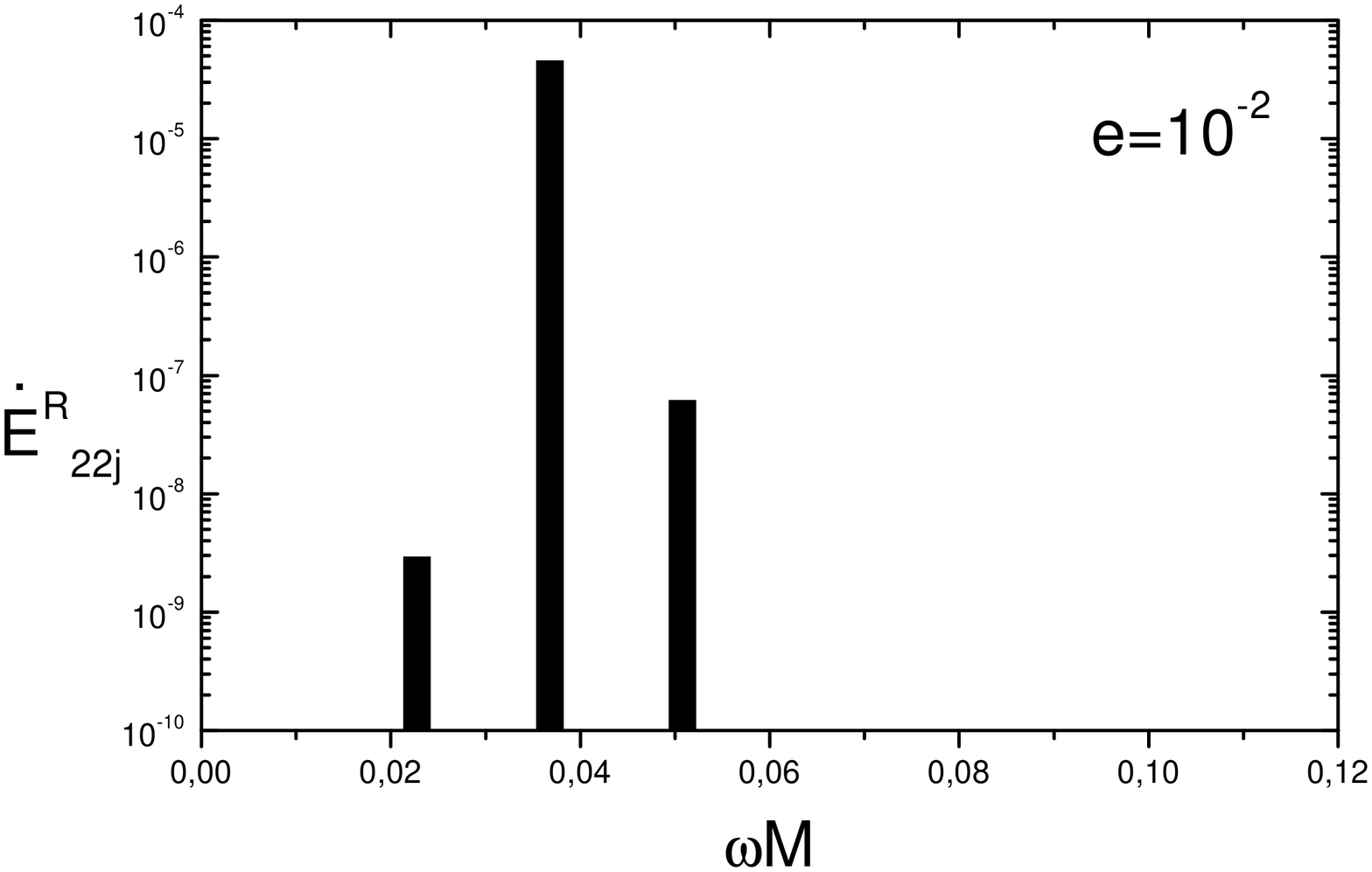}
}}
\vskip 8pt
\centerline{\mbox{      
\epsfxsize=10cm \epsfbox{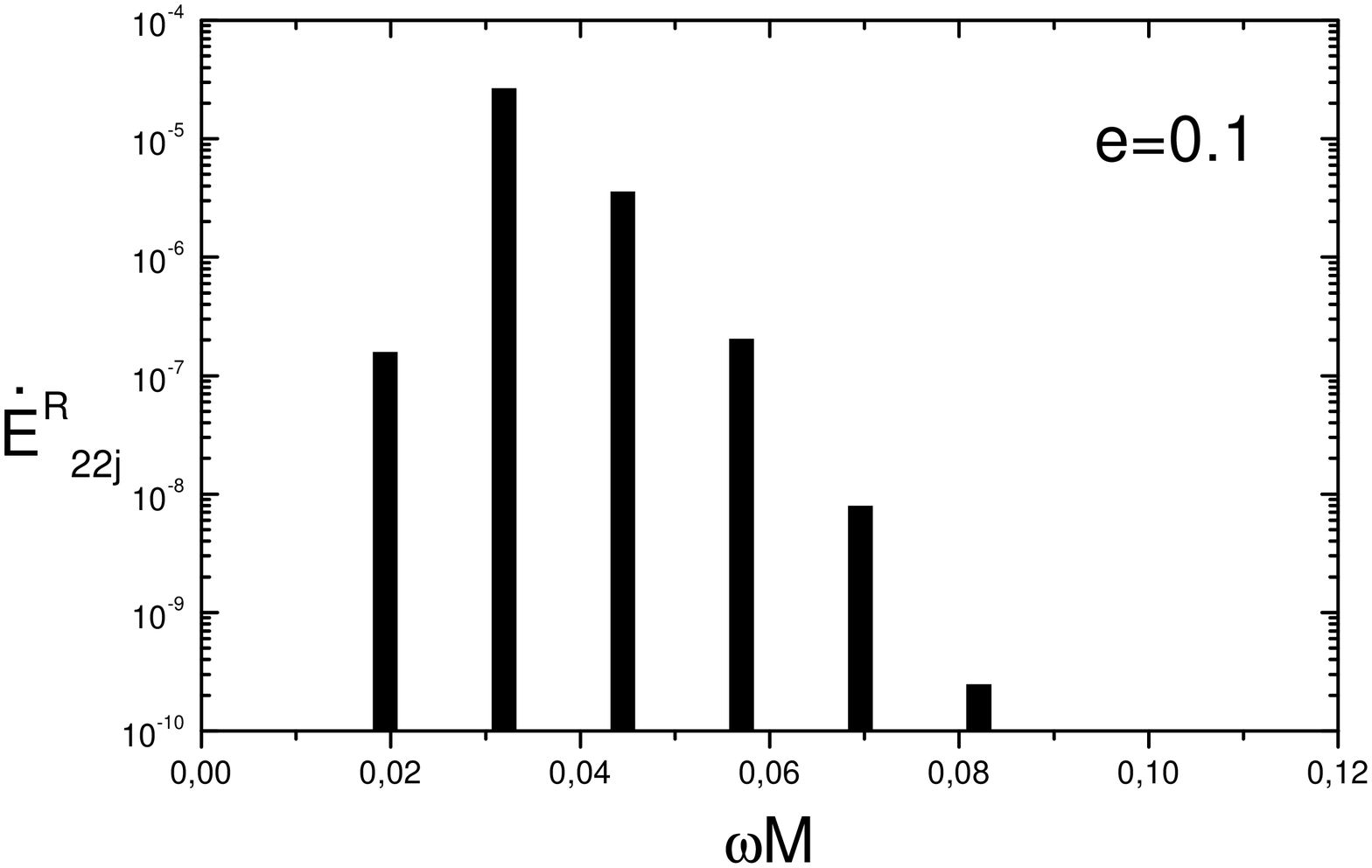}
\epsfxsize=10cm \epsfbox{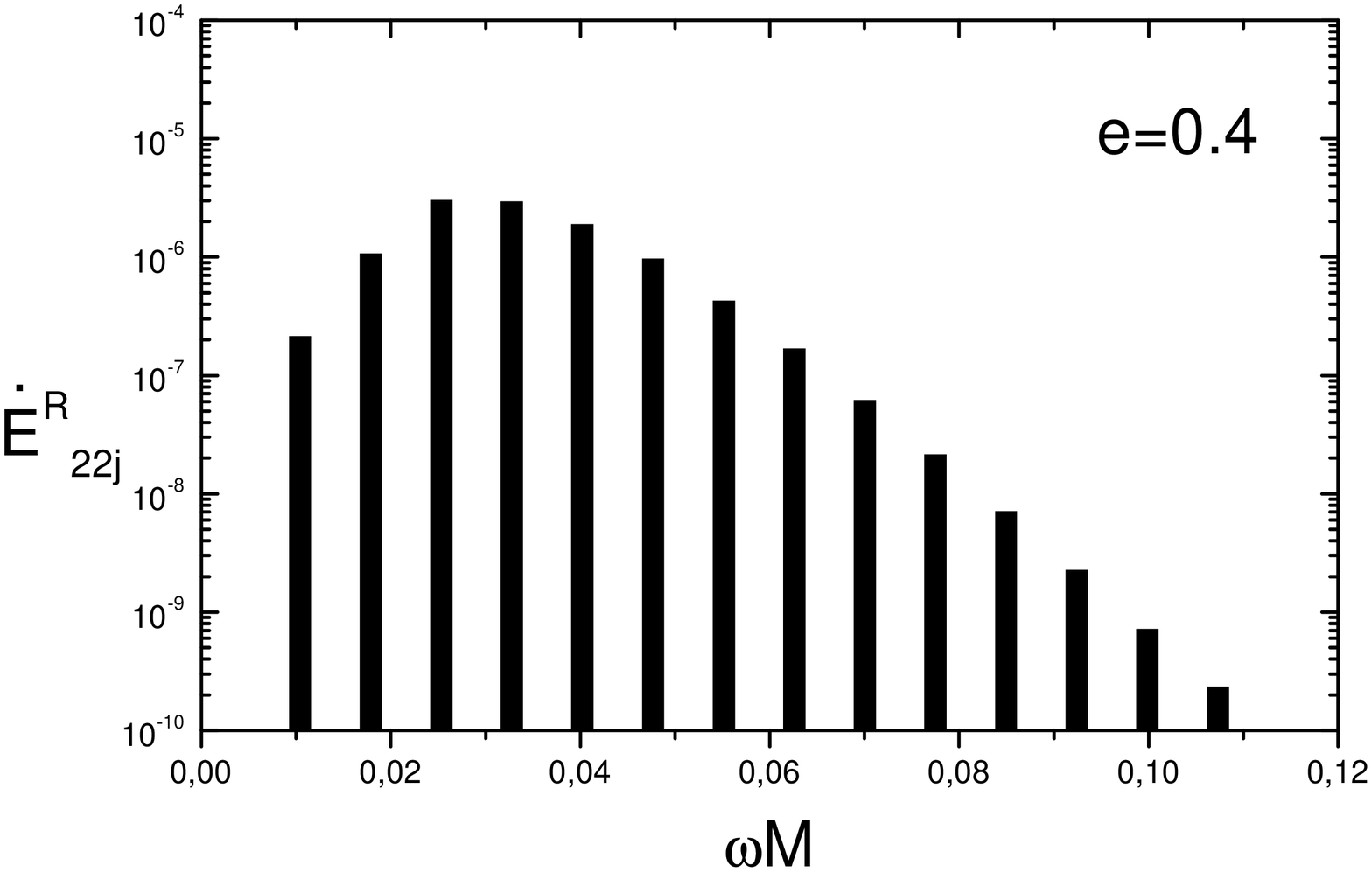}
}}
\vskip 10pt
\caption{
In this figure we show how the spectral content of the gravitational emission 
changes as a function of the eccentricity, plotting
the spectral lines \op \dot E^R_{l m j }\cl 
as a function of the dimensionless frequency \op M\omega_{m j}, \cl
for $l=m=2$  and for the same value of the periastron $r_P=3~R_s$. 
As the eccentricity increases, the location of the highest 
line shifts slightly towards a lower frequency, and
higher order harmonics become more relevant.
}\label{ecc}
\end{figure}

In Figure \ref{ecc} we show how
the energy output obtained by the relativistic approach varies as a function of
the eccentricity of the orbit; we consider four cases,
\op e= 10^{-3}, 10^{-2}, 0.1, 0.4. \cl 
All orbits have the same periastron  ($r_P=3 R_s$),
and the plots are given for  $l=m=2$, since  this is the
dominant contribution to the emitted radiation.
In  the zero eccentricity limit,
the whole power is concentrated in the harmonic with $j=0$, corresponding to a 
frequency $\omega_{circ}=2\f{\Delta \varphi}{\Delta t}=2 \omega_k$,
where  $\omega_k$ is the keplerian orbital frequency; 
as the eccentricity increases, the frequency of the highest line
slightly decreases, and higher order harmonics become significant.

\begin{figure}[htbp]
\begin{center}
\leavevmode
\epsfxsize=16cm \epsfbox{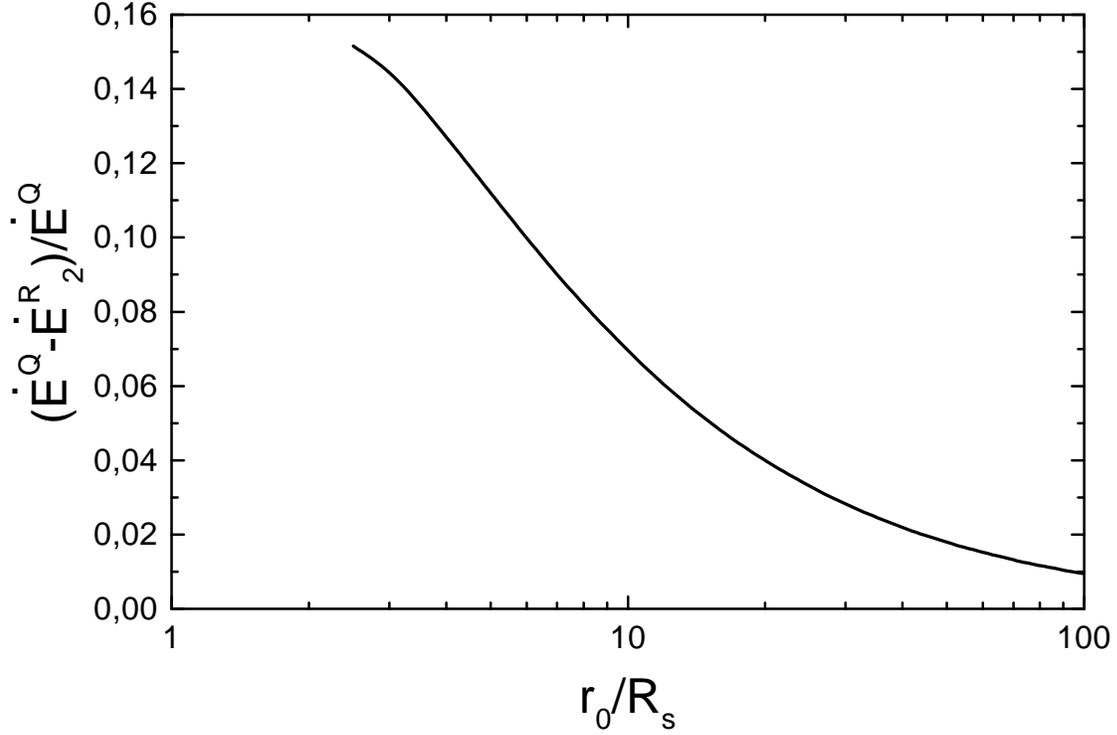}
\end{center}
\caption{
The relative difference between the total power $\dot{E}^{Q}$
computed by the hybrid quadrupole approximation,
and the total power emitted in the $l=2$ multipole,  
$\dot{E}^{(R)}_2$, computed by the relativistic approach,
is plotted for circular orbits as a function  of the radius $r_0$
(given in units of the stellar radius).
When the point mass moves on an orbit far from the star
the two approaches give the same result; 
the quadrupole emission becomes significantly larger than $\dot{E}^{(R)}_2$ 
for \op r_0 < 10~R_S.\cl
}\label{relerror}
\end{figure}

We have compared  the total power   computed by the quadrupole formalism
\be
\label{totpowquad}
\dot E^Q = \sum_{j=-\infty}^\infty~\sum_{m=-2,0,2} \dot E^Q_{mj}
\ee
with  that emitted in the $l=2$ multipole, \op \dot E^R_2,\cl 
defined in eq. (\ref{totpow}) and computed by the perturbative approach,
for different values of the eccentricity and
of the periastron. We find that,  in general,
\op \dot E^R_2\cl is sistematically smaller than
\op \dot E^Q.\cl
The amount of emitted radiation affects the orbital evolution of the system
and the shape of the gravitational signal, in particular
during the latest phases of coalescence, where the orbit is already circularized.
To understand the relevance of this effect, which may be important 
for the detection of these signals by the ground based interferometers VIRGO and LIGO,
we have computed the relative difference  
\be
\label{reldiff}
\dps{
\frac{\dot E^R_2-  \dot E^Q}{ \dot E^Q}
}
\ee
for circular orbits, as a function of the orbital radius,  $r_0$.
The results are shown in Figure \ref{relerror}. 
For large values of the radius the relative difference tends, as expected, to zero;
at a distance of 10 stellar radii it is about 7 \%, and it becomes 
greater than 14 \% when the two stars are \op 3~R_s\cl apart.

In order to  check the correctness of our results, we have
repeated the calculation  by a different approach,  integrating the 
inhomogeneous Zerilli and the Regge-Wheeler equations for the same orbits. 
The results agree to the  round-off error.

The situation changes if the point mass moves on an orbit ``resonant" with 
a mode of the star, which means the following. 
For the model of star we are considering, the lowest
frequency mode is the fundamental one, whose frequency is \op\omega_f M=0.12034.\cl
To excite this mode the mass should move on an orbit 
such that the frequency of one of the spectral lines of the quadrupole emission, 
\op \omega_{m j},\cl with \op m=-2,0,2,\cl
is very close to, or coincides with \op\omega_f.\cl
We find  that, as the quadrupole spectral line frequency
\op \omega_{m j}\cl approaches \op\omega_f\cl
for some value of $m$ and $j$,  
the amplitude of the emitted radiation,
computed in the perturbative approach,
increases. This suggests that the excitation
mechanism could be seen as a resonant scattering of the gravitational wave
emitted by the system in the orbital motion (the quadrupole wave) 
on the potential barrier generated  by the perturbed star.
Indeed, the discrete nature of the power spectrum emitted in the quasi 
periodic motion of the point mass suggests an analogy with an atomic laser:
in this picture, the atomic energy levels correspond to the quasi-normal mode 
frequencies, and the quadrupole radiation frequencies to the energy 
of the electromagnetic radiation exciting them.

\begin{figure}[htbp]
\begin{center}
\leavevmode
\epsfxsize=16cm \epsfbox{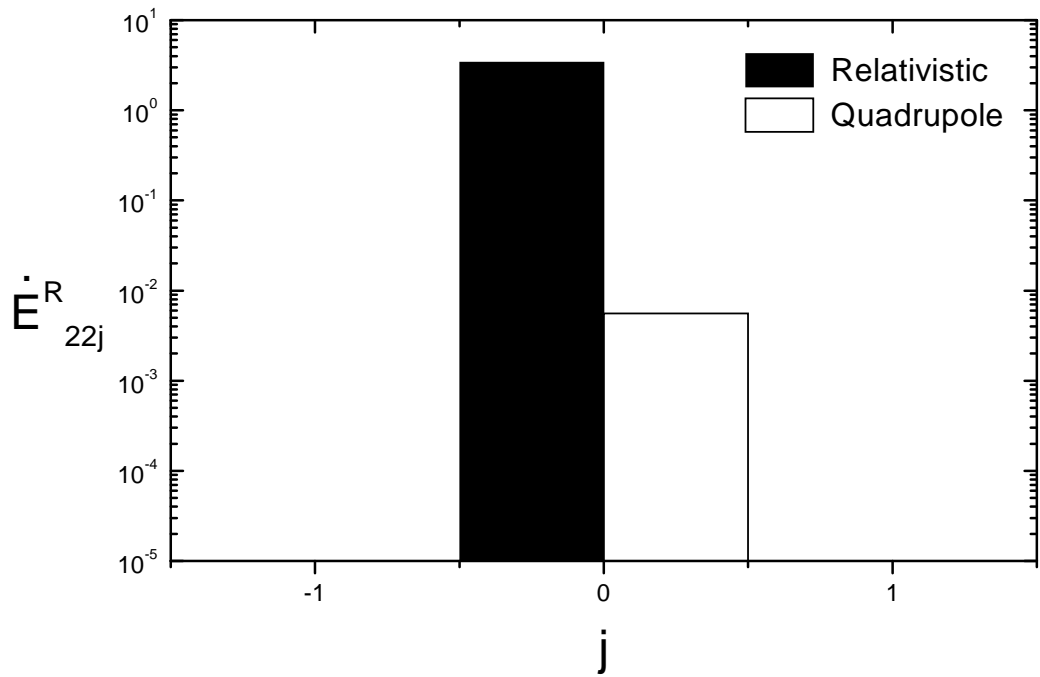}
\end{center}
\caption{
The spectral line emitted by the system when 
the point mass moves on a  close circular orbit 
is compared to the same quantity computed by the 
hybrid quadrupole formalism. The  relativistic line (in black) is much larger
than the quadrupole  one (in white), because the frequency of
the quadrupole line   coincides with that of the fundamental
mode of the star, \op\omega_f,\cl and a mechanism of 
resonant excitation occurs (see text).
}\label{fmode}
\end{figure}

In order to excite the fundamental mode of our star,
the two bodies must be very close, and therefore 
it is reasonable to assume that the orbit is circular and that 
\op \omega_{m j}\cl reduces to \op \omega_{circ}\cl defined above.
To show how efficient this resonant mechanism could be, in 
Figure \ref{fmode} we plot the energy output  \op \dot E^R_{2 2 j},\cl
and the corresponding quadrupole energy \op\dot E^Q_{2 j}\cl
for a point mass moving on an  orbit such that \op \omega_{circ}=\omega_f\cl
\op (r_0=1.37417~R_s).\cl
From this figure we see that the situation changes
dramatically with respect to the non-resonant case:  
the energy emitted in the relativistic calculation
is about $600$ times larger than that 
computed  by the quadrupole approximation. 
Whether the fundamental mode could be excited during the coalescence of 
neutron star binary systems is, however, questionable, and will be 
discussed in the concluding remarks.

It should be mentioned that the $f$-mode excitation  by a
particle in  circular orbit around a star was  studied
also by  Kojima  \cite{kojima}.
In his paper he only considered circular orbits and polar perturbations
with $l=2,~ m=\pm 2$.
We find the same qualitative behaviour,  although  with
minor differences of the order of 10\% in the  wave amplitude.
We are confident on the correctness of our results
since, as mentioned above, we got them using two completely different
formalisms, one based on the Regge-Wheeler, the other on the Newman-Penrose
approach.

\begin{figure}[htbp]
\begin{center}
\leavevmode
\epsfxsize=16cm \epsfbox{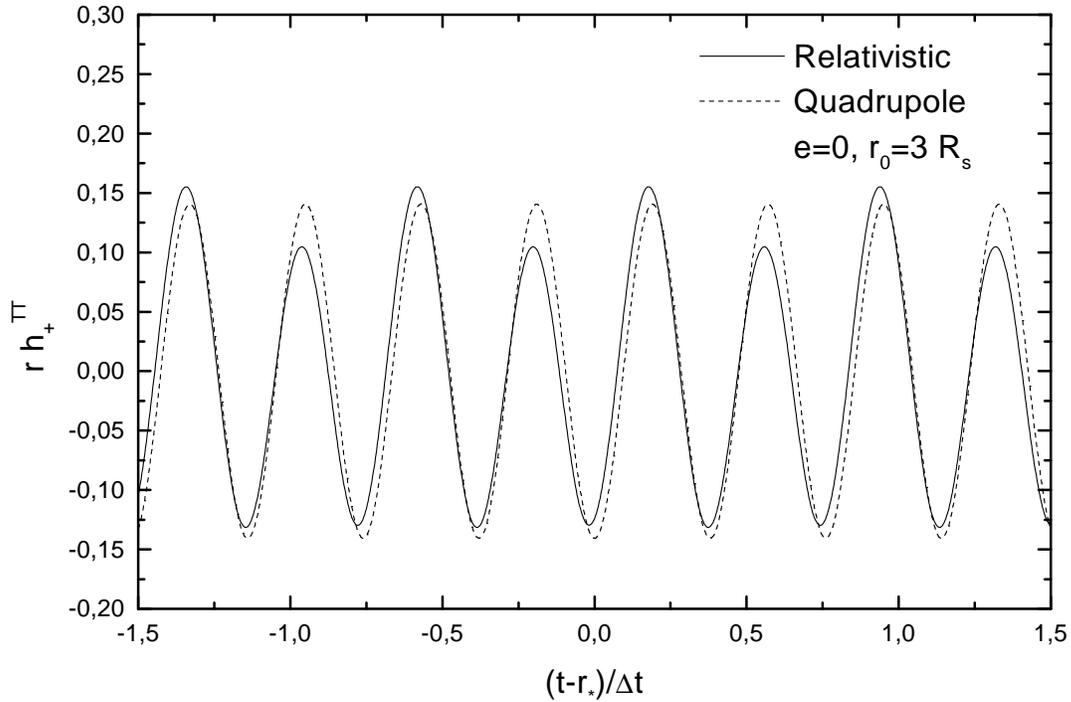}
\end{center}
\caption{
The  \op h^{TT}_+\cl component of the gravitational wave emitted 
when the point mass moves on a circular orbit with \op r_0=3~R_s,\cl  
is plotted versus the retarded time in units of the orbital period.
Since we  assume that the observer is on the equatorial plane, 
the \op h^{TT}_\times\cl component vanishes.
In order to compare the relativistic waveform (continuous line) 
with the  waveform computed by the hybrid quadrupole approach (dashed line), 
only the $l=2$ component of the relativistic signal is shown.
The difference between the two  signals is basically due to the
$m=1$ contribution of the axial perturbations to the relativistic waveform
(see text).
}\label{waveformcirc}
\end{figure}

In Figure \ref{waveformcirc}, we show 
the $+$ polarization of the waveform - the $\times$ polarization
is zero because we assume the observer is on the equatorial plane - for
a circular orbit with radius
\op r_0= 3 R_s, \cl  and for \op l=2.\cl
The gravitational waveform obtained in the quadrupole
approximation is also shown (dashed lines) for comparison. 
There are two remarkable effects. 
The first is that the $m=1$ axial contribution to the relativistic waveform, 
which is usually ignored, is not negligible.
Indeed it induces a beating of the axial 
and polar frequencies, clearly seen in the figure.
The second  effect is that the average of the amplitudes  of
the positive (or of the negative) peaks in the relativistic waveform,
is smaller than that of the quadrupole waveform. 
This is related to the fact that the $l=2$
relativistic energy output is systematically smaller than that of the quadrupole
as we get close to the star (see Figure 3). 

\begin{figure}[htbp]
\begin{center}
\leavevmode
\epsfxsize=16cm \epsfbox{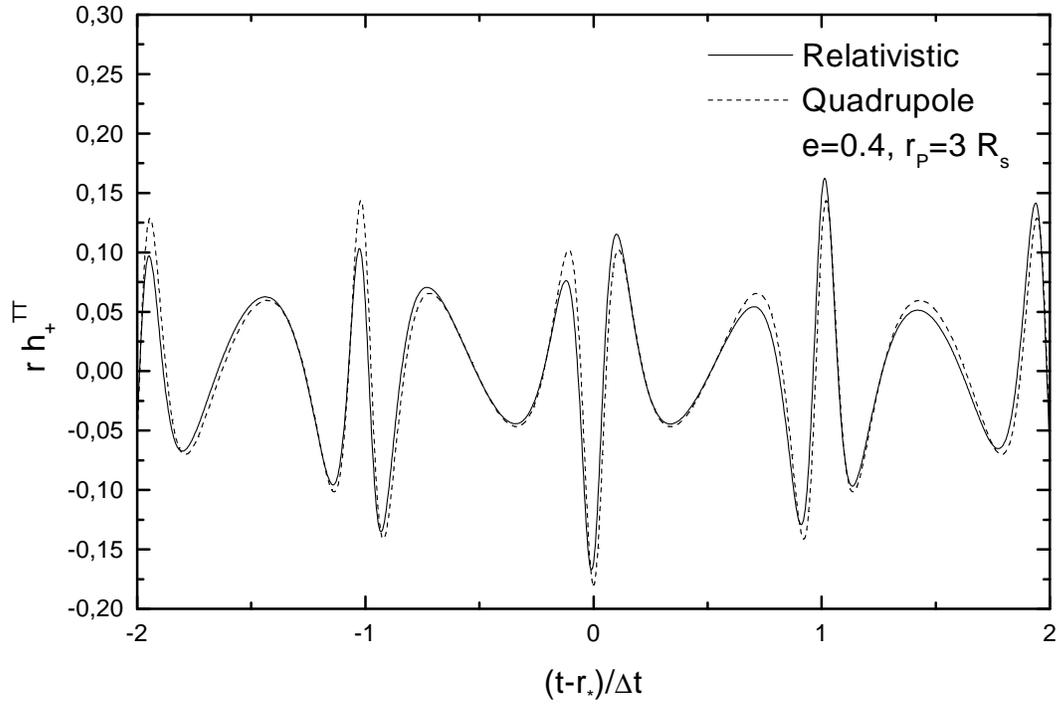}
\end{center}
\caption{
As in figure 6, we plot 
the  \op h^{TT}_+\cl component of the gravitational wave emitted 
when the point mass moves on an eccentric orbit with \op e=0.4\cl  
and \op r_P= 3 R_s.\cl
The structure of the waveform is now much more complicated,
but the beating of the
frequencies induced by the $m=1$ axial contribution is still present.
}\label{waveformellipt}
\end{figure}

A case with large eccentricity ($e=0.4$) is shown in Figure \ref{waveformellipt}. 
The structure of the waveforms is now much more complicated. However,
both effects seen in the circular case are still present. The beating of the
axial and polar frequencies produces similar changes in the maxima  
of the wave amplitude, i.e   about 10 \% in both cases. 
We have found that the relative contribution
between the axial and polar emission \op \dot E^{R~ax}_l/ \dot E^{R~pol}_l\cl
is quite independent of eccentricity and decreases approximately as
 $\approx 1/p$. Thus, it becomes negligible at large
distances but it might be significant in the late stages of 
the inspiralling.

\section{Concluding remarks}
In this paper we have studied the gravitational emission of a binary 
system composed of a star and a point mass orbiting around it,
by using a perturbative approach. The results have been compared
with the orbital emission  computed by the quadrupole formalism,
which assumes that both objects are pointlike, thus neglecting the 
fact  that the star is an extended body with an internal structure,
and that  the dynamical  evolution of the gravitational field  couples 
to the thermodynamical evolution of the star.
Of course the perturbative approach also is  quite a  crude approximation 
of a realistic binary system, since one of the two stars is still considered
as a point mass. However, it allows to treat at least one star
in an  exact manner, since its internal structure and its gravitational
field are exact solutions of  the equations of hydrostatic equilibrium. The 
interaction with the companion is treated  as a perturbation, 
and evaluated by linearizing  the Einstein equations coupled with the hydrodynamical
equations.

In comparing the outcome of the two approaches, we find the following effects:\\
1) The total relativistic power \op\dot E^R_{2}\cl
emitted in the $l=2$ multipole
is smaller than that computed by the quadrupole approach, \op\dot E^Q.\cl\\
2) The waveforms have a different shape, due to the fact that the $m=1$ axial
contribution, which is absent in the quadrupole scheme,
produces a beating of the axial and polar frequencies.\\
3) If the point mass is allowed to get sufficiently close as to excite the fundamental 
mode of the star, the amplitude of the wave computed by the relativistic approach
significantly increases with respect to the quadrupole prediction.
About this point, it should be noted that
the frequency of the fundamental mode of neutron stars, \op\omega_f,\cl
is expected to be of the order
of \op \sim 2-3\cl kHz, depending on the equation of state prevailing in the interior. 
This frequency is too high to be excited, since the coalescing system would reach 
the ISCO (Innermost Stable Circular Orbit) and merge, before the resonant orbit 
is reached. However, \op\omega_f\cl
is affected by rotation, and for fast rotating
stars it may become small enough to be ``excitable".  In that case the emitted 
radiation would  probably be enhanced by the rotation, and this is an 
interesting effect that has never been studied in a relativistic framework.

In this paper we have considered a very simple model of star described by a 
polytropic equation of state. More realistic models of neutron stars 
allow the existence of other classes of modes at lower frequency with respect to 
\op\omega_f.\cl For instance, the $g$-modes may lay in a frequency range 
of about $\sim 100-400$ Hz, which is the region where the ground-based
interferometers are more sensitive. If the resonant excitation of these modes
is efficient, the signal emitted during coalescence may suddenly change
when the resonant orbit is approached, introducing a new feature in 
the expected waveform.

Besides this, several other issues remain to be clarified.
The waveforms we produce by the 
relativistic approach are different from those  computed 
by the hybrid quadrupole approach.
Although this is  more accurate than the newtonian 
quadrupole approach, because it assumes that the pointlike mass moves on a geodesic 
of the unperturbed spacetime, we
are aware of the fact that by the PPN formalism (see e.g. 
\cite{damouriyersathya})  it is possible 
to refine the trajectory of the particle especially near coalescence, 
and have a more accurate evaluation of the radiation emitted because of the 
orbital motion.
Thus the question is: is the difference between the signal we compute and 
the most accurate estimate of the signal which is provided by the PPN formalism
still significant? We believe the answer is positive, because the difference
between the relativistic and the quadrupole signals we find can be attributed 
to the role played  by the internal structure of the star, and to the way in which
the gravitational field couples with the fluid.
However, this question has to be answered by a direct comparison.

In order to produce waveforms that can be used as templates in the data
analysis of gravitational wave experiments, radiation reaction effects have to 
be considered.  We are working to include 
these effects in our scheme following refs. 
\cite{CutlerKennefickPoisson,BertiFerrari},
and to evaluate how the evolution of the system changes with respect to the
traditional picture. 
Important questions that need to be answered to 
construct a matched filter  and to extract
the chirp mass of the coalescing system are: i) 
what is the number of cycles the gravitational signal does in the bandwidth of
the interferometers, ii)  how the amplitude changes in time,
iii) how much these effects depend
on the equation of state of dense matter.
All these issues will be considered in subsequent papers.

\acknowledgments
This work has been supported by the EU Programme 'Improving the Human
Research Potential and the Socio-Economic Knowledge Base' (Research
Training Network Contract HPRN-CT-2000-00137).

\appendix
\section{The source term of the BPT equation}
In this appendix we discuss the procedure to find the wave 
amplitude \op A_{l m}(\o)\cl (cfr. Eq.\ref{ampli1})
\be
\label{ampli1a}
A_{l m}(\o)=
-{1\over W_{l m}(\o)}~
{2\pi\over\Delta t} \sum_{j=-\infty}^{+\infty}
\delta\left(\omega-\omega_{m j}\right)
\int_{R_s}^{\infty}~ \frac{dr'}{\Delta^2}~\Psi^{~1}_{l m}(\o,r')~ 
T_{0~l m}(\o,r'),
\ee
where
\be
\label{ttildenew}
T_{0~l m}(r,\o)=-2\sqrt{n(n+1)}r^4 T_{0(n)(n)}
-2\sqrt{n}\Delta\Lambda_+{r^5\over\Delta} T_{0(n)(\bar m)}
-{\Delta\over 2r}\Lambda_+{r^6\over\Delta}\Lambda_+r T_{0(\bar m)
(\bar m)}\,.
\ee
The tetrad components 
of the stress energy tensor of the pointlike mass  are
\beqn
T_{0(n)(n) l m}(\omega,r)&=&
\left[{1\over r^2}\left(\gamma+{E^2\over\gamma}\right) \cos(\o t-m\varphi)+
\ii~{2E\over r^2}\sin(\o t-m\varphi)\right]~_0S^*_{l m}(\pi/2,0)~,\\
T_{0(n)(\bar{m}) l m}(\omega,r)&=&
\frac{\sqrt{2}L_z}{r^3}
\left[- \sin(\o t-m\varphi)+
{E\over\gamma }~ \cos(\o t-m\varphi)\right]~_{-1}S^*_{l m}(\pi/2,0)~,\\
T_{0(\bar{m})(\bar{m}) l m}(\omega,r)&=&-\frac{2L_z^2}{\gamma r^4}
\cos(\o t-m\varphi)~_{-2}S^*_{l m}(\pi/2,0)~,
\eeqn
where \op r, t(r)$ and $\varphi(r)\cl refer to the point mass trajectory.
From the expression of \op T_{0~l m}(r,\o)\cl we see that the last two terms 
contain the differential operator \op\Lambda_+\cl applied to a function of $r;$
when we  evaluate the integral (\ref{ampli1a})
these terms can be integrated by parts
defining the operators \op \hat\Lambda_\pm=\frac{r^2}{\Delta} \Lambda_\pm,
\cl where 
\op \Lambda_\pm={d\over dr_*}\pm i\omega,\cl
and using  the property
\be
\int_{R_s}^{\infty}drf(r)\hat \Lambda_+g(r)=
-\int_{R_s}^{\infty}drg(r)\hat \Lambda_-f(r)
\ee
which holds if, as always in our case, $f(r)$ or $g(r)$ vanishes at the extrema of the 
integration.  After applying this procedure, and replacing 
the expressions of the \op T_{0(p)(q) l m},\cl
the wave amplitude can  be written as
\be
\label{ampli1b}
A_{l m}(\o)=
-\frac{1}{W_{l m}(\o)}~
\frac{2\pi}{\Delta t} \sum_{j=-\infty}^{+\infty}
\delta\left(\omega-\omega_{m j}\right)
\int_{R_s}^{\infty}~ dr\left[I_0+I_{-1}+I_{-2}\right],
\ee
where
\beqn
I_0 &=&
-2\sqrt{n(n+1)}\frac{r^4}{\Delta^2} T_{0(n)(n) l m} ~\Psi^{~1}_{l m}
\\
&=&-2\sqrt{n(n+1)}\frac{r^2}{\Delta^2}
\left[\frac{(\gamma^2+E^2)}{\gamma}\cos(\o t-m\varphi)+
2\ii E\sin(\o t-m\varphi) \right] 
~\Psi^{~1}_{l m}(\o,r) ~_0S^*_{l m}(\pi/2,0)
\\
I_{-1} &=&
-2\sqrt{n}\frac{~\Psi^{~1}_{l m}}{\Delta}\Lambda_+
\frac{r^5}{\Delta}T_{0 (n)(\bar{m}) l m}
\\
&=&-\frac{2\sqrt{2n}L_z}{\Delta}
\left[\sin(\o t-m\varphi)-i\frac{E}{\gamma}\cos(\o t-m\varphi)\right]
\left[
~\Psi^{~1~'}_{l m}(\o,r) \right.
\\
&-&\left.
\left(\frac{2}{r}+i\frac{\omega r^2}{\Delta}\right)
~\Psi^{~1}_{l m}(\o,r)\right]
~_1S^*_{l m}(\pi/2,0)  
\\
I_{-2}&=&
-\frac{\Psi^{~1}_{l m}}{2r\Delta}\Lambda_+\frac{r^6}{\Delta}\Lambda_+r
T_{0 (\bar{m})(\bar{m}) l m}
\\
&=&-\frac{L_z^2}{\gamma r^4}\cos(\o t-m\varphi)\left[
r^2 \Psi^{~1~''}_{l m}(\o,r)
-2\left(r+i\omega\frac{r^4}{\Delta}\right)
\Psi^{~1~'}_{l m}(\o,r)\right.
\\
&+&\left.
\left(2i\omega\frac{r^4}{\Delta^2}(r-M)-
{\omega^2 r^6\over \Delta^2}\right)
\Psi^{~1}_{l m}(\o,r)
\right]
~_2S^*_{l m}(\pi/2,0) .
\eeqn
Since the integrand of Eq. (\ref{ampli1b}) diverges at the turning points where
\op\gamma=0,\cl it is convenient to perform the numerical integration using a
different integration variable, defined by the following equation
\be
r(\chi)=\f{pM}{1+e\cos\chi}~,
\ee
where  $\chi$ ranges from $0$ to $2\pi$ in a whole orbit.
In terms of  $\chi$ the equations of motion  become:
\beq
\f{dt}{d\chi}&=&\f{p^2M(p-2-2e)^{1/2}(p-2+2e)^{1/2}}
{(p-2-2e\cos\chi)(1+e\cos\chi)^2(p-6-2e\cos\chi)^{1/2}} \\
\f{d\tau}{d\chi}&=&\f{p^{3/2}M(p-3-e^2)^{1/2}}
{(1+e\cos\chi)^2(p-6-2e\cos\chi)^{1/2}} \\
\f{d\varphi}{d\chi}&=&\f{p^{1/2}}{(p-6-2e\cos\chi)^{1/2}}.
\eeq



\section{Fourier transform of the quadrupole wave}

In this appendix we explicitely compute the Fourier
transform of the metric components  of the
wave emitted by the pointlike particle because of its orbital motion,
given in Eq. (\ref{htp}), and discuss the symmetry properties of the
corresponding spectrum.
The orbit is  quasi-periodic in  $\varphi$,
but exactly periodic in ($r, \dot{r}, \ddot{r}, \dot{\varphi}, \ddot{\varphi}$)
so that we can 
decompose the Fourier transform  as  a sum over periods
\beq
h(\omega)&=&\f{1}{2\pi}\int_{-\infty}^{\infty}h(t)e^{i\omega t}dt=
\\
\nn
&=&\f{1}{2\pi}\sum_{n=-\infty}^{\infty}\int_{(n-1/2)\Delta
t}^{(n+1/2)\Delta t}h
(t_n=t_0+n\Delta t,\varphi_n=\varphi_0+n\Delta \varphi)e^{i\omega t_n}dt_n,
\eeq
where  $\left(t_0(r),\varphi_0(r)\right)$ indicates
the branch of the trajectory which starts at the periastron 
($r=r_P$) at $t=0$ and ends at the apoastron ($r=r_A$) 
at $t=\f{\Delta t}{2};$  $\left(-t_0(r),-\varphi_0(r)\right)$ 
indicates the ``mirror'' branch starting at $r_A$ and ending at  $r_P$, 
with $t\in [-\f{\Delta t}{2},0]$.
An inspection of Eqs.  (\ref{htp}) and (\ref{qxx})
shows that the integrals are essentially
of three types: 1) those containing $\delta$, say $h_1(t)$, which 
do not depend on $\varphi$ and therefore are exactly periodic; 
2) those that contain a periodic term, say $h_2(t)$, times $\sin2\varphi$, 
and
3) those that contain a periodic term, say $h_3(t)$, times $\cos2\varphi$.
These integrals can be developed in the following way:

\beqn
1)&&\f{1}{2\pi}\sum_{n=-\infty}^{\infty}
\int_{(n-1/2)\Delta t}^{(n+1/2)\Delta t}h_{1}(t_n)e^{i\omega t_n}dt_n
=\f{1}{2\pi}\int_{-\Delta t/2}^{\Delta t/2}h_{1}(t_0)e^{i\omega
t_0}\sum_{n= -\infty}^{\infty}e^{i\omega n\Delta t}dt_0=\\
&=& \sum_{j=-\infty}^{\infty}\delta\left(\omega-\omega_{j0}\right) 
\left[\f{1}{\Delta t}\int_{-\Delta t/2}^{\Delta t/2}h_{1}(t_0)e^{i\omega
t_0}dt_0\right].
\eeqn
To obtain this result we have used Eq. (\ref{deltas}) and the definition 
(\ref{freq}) of $\omega_{m j}$.
Similarly we have:
\beqn
2)&&\f{1}{2\pi} \sum_{n=-\infty}^{\infty}
\int_{(n-1/2)\Delta t}^{(n+1/2)\Delta t}h_{2}(t_n)
\sin2\varphi_n e^{i\omega t_n}dt_n
\\
&=&
\sum_{j=-\infty}^{\infty}
\left\{
\delta\left(\omega-\omega_{j-2}\right)  
\left[\f{1}{\Delta t} 
\int_{-\Delta t/2}^{\Delta t/2} h_{2}(t_0) 
\f{e^{i(\omega t_0+2\varphi_0)} }{2i} dt_0\right]
\right.
\\
&+&\left.
\delta\left(\omega-\omega_{j2}\right)  
\left[-\f{1}{\Delta t}
\int_{-\Delta t/2}^{\Delta t/2}h_{2}(t_0)
\f{e^{i(\omega t_0-2\varphi_0)}}{2i}dt_0\right]
\right\}.
\eeqn

\beqn
3)&&\f{1}{2\pi}
\sum_{n=-\infty}^{\infty}
\int_{(n-1/2)\Delta t}^{(n+1/2)\Delta t}h_{3}(t_n) 
\cos2\varphi_n e^{i\omega t_n}dt_n
\\
&=&
\sum_{j=-\infty}^{\infty}
\left\{
\delta\left(\omega-\omega_{j-2}\right) 
\left[\f{1}{\Delta t} \int_{-\Delta t/2}^{\Delta t/2} h_{3}(t_0) \f{
e^{i(\omega t_0+2\varphi_0)} }{2} dt_0\right]
\right.
\\
&+&\left.
\delta\left(\omega-\omega_{j2}\right) 
\left[\f{1}{\Delta t}\int_{-\Delta t/2}^{\Delta
t/2}h_{3}(t_0)\f{e^{i(\omega t _0-2\varphi_0)}}{2}dt_0\right]
\right\}.
\eeqn
By using this procedure we find that the wave components
can be written as
\beq
h^{TT}_{\theta \theta}(\omega, r, \theta, \phi)&=&
\sum_{j=-\infty}^{\infty} \sum_{m=-2,0,2}
\delta(\omega-\omega_{m j})~H_{\theta \theta}(\omega_{m j},\theta,\phi)
\\\nn
h^{TT}_{\theta \phi}(\omega, r, \theta, \phi)&=&
- \sum_{j=-\infty}^{\infty}\sum_{m=-2,2}   
\delta(\omega-\omega_{m j})~H_{\theta \phi}(\omega_{m j},\theta,\phi)
\eeq
where 
\beqn
H_{\theta \theta}(\omega_{2 j},\theta,\phi)
&=&\f{m_0 }{ r}\left[
\left( {\cal I}^\alpha_{2 j} + {\cal I}^\beta_{2 j} \right)
(1+\cos^2\theta)(\cos 2\phi+i\sin 2\phi) \right]\\
H_{\theta \theta}(\omega_{-2 j},\theta,\phi)
&=&\f{m_0 }{ r}\left[
\left( {\cal I}^\alpha_{-2 j} - {\cal I}^\beta_{-2 j} \right)
(1+\cos^2\theta)(\cos2\phi-i\sin2\phi)
\right]\\
H_{\theta \theta}(\omega_{0 j},\theta,\phi)
&=& \f{3m_0 }{ r}{\cal K}^\delta_j(\cos^2\theta-1)\\
H_{\theta \phi}(\omega_{2 j},\theta,\phi)
&=&
\f{2m_0 }{ r}\cos\theta \left[
\left( {\cal I}^\alpha_{2 j} + {\cal I}^\beta_{2 j} \right)
(\sin 2\phi-i\cos 2\phi)
\right]\\
H_{\theta \phi}(\omega_{-2 j},\theta,\phi)
&=&
\f{2m_0 }{ r}\cos\theta
\left[\left( {\cal I}^\alpha_{-2 j} - {\cal I}^\beta_{-2 j} \right)
(\sin 2\phi+i\cos 2\phi)
\right]
\eeqn
and
\beqn
{\cal I}^\alpha_{\pm 2 j}&=&
\f{1}{\Delta t}\int_{0}^{\Delta t/2}\alpha \cos(\omega_{j \pm 2} t_0 \mp 2\varphi_0)
dt_0
\\
{\cal I}^\beta_{\pm 2 j}&=&
\f{1}{\Delta t}\int_{0}^{\Delta t/2}\beta \sin(\omega_{j \pm 2} t_0 \mp 2\varphi_0)
dt_0
\\
{\cal K}^\delta_j&=&
\f{1}{\Delta t}\int_{0}^{\Delta t/2}\delta \cos(\omega_{j0} t_0) dt_0
\eeqn

In terms of these integrals, the various contributions to the average 
power radiated appearing in formula (\ref{quadflux}) are given by
\beq
\dot E^{Q}_{2 j}&=&\f{4m_0^2}{5}
(\omega_{2 j})^2
\left({\cal I}^\alpha_{2 j}+{\cal I}^\beta_{2 j}\right)^2,
\nn \\
\dot E^{Q}_{-2 j}&=&\f{4m_0^2}{5}
(\omega_{-2 j})^2
\left({\cal I}^\alpha_{-2 j}-{\cal I}^\beta_{-2 j}\right)^2,
\nn \\
\dot E^{Q}_{0 j}&=&\f{6m_0^2}{5}
\omega_{0 j}^2 \left({\cal K}^\delta_j\right)^2.
\nn
\eeq
Using Eq. (\ref{symmOmega}) and the definitions of the various integrals, 
it is straightforward to prove that:
\beq
&&{\cal I}^\alpha_{-2 -j}={\cal I}^\alpha_{2 j},\qquad 
{\cal I}^\alpha_{2 -j}={\cal I}^\alpha_{-2 j}, \nn\\
&&{\cal I}^\beta_{-2-j}=-{\cal I}^\beta_{2 j},\qquad
{\cal I}^\beta_{2 -j}=-{\cal I}^\beta_{-2 j}, \nn\\
&&{\cal K}^\delta_{-j}={\cal K}^\delta_{j}.
\label{quadsymmetry}
\eeq
Finally, from  Eqs. (\ref{symmOmega}) and (\ref{quadsymmetry}) we find 
\beq
\dot E^{Q}_{-m-j}=\dot E^{Q}_{m j} \qquad (m=2,0,-2)
\eeq
that are similar to the properties
(\ref{energyparity}) for the relativistic energy flux.

Thus the quadrupole power spectrum has essentially the same frequency content
and the same symmetry properties as the the relativistic spectrum
with $l=2$ and $m=-2,0,2$, except for the $m=1$ contribution, which is missing.

\begin{table}
\centering
\caption{Ratio of the power in the $l$-th multipole, \op \dot E^R_l,\cl to the
power in $l=2$ multipole for two circular orbits of radius $r_0=3 R_s$
and $r_0=10 R_s,$ respectively.
}
\vskip 12pt
\begin{tabular}{@{}lrr@{}}
\multicolumn{1}{c}{} & $\dot E^R_l/\dot E^R_2 $ & \\
\hline
\multicolumn{1}{c}{$l$}&$3 R_s$  &$10 R_s$  \\
\hline
3 & $8.6\cdot 10^{-2}$& $2.7\cdot 10^{-2}$ \\
4 & $9.2\cdot 10^{-3}$& $9.2\cdot 10^{-4}$ \\
5 & $1.0\cdot 10^{-3}$& $3.3\cdot 10^{-5}$ \\
6 & $1.2\cdot 10^{-4}$& $1.2\cdot 10^{-6}$ \\
\end{tabular}
\end{table}

\newpage

\newpage

\newpage

\begin{references}


\bibitem{nonlinear}
K. Oohara, T.Nakamura,  Prog. Theor.  Phys. Suppl. {\bf 136 }, 270, (1999); 
M. Ruffert, H.T. Janka, Prog. Theor.  Phys. Suppl. {\bf 136 }, 287, (1999) ;
F.A. Rasio, S.L. Shapiro, Class. Quant. Grav. {\bf 16}, 1, (1999);
T.W. Baumgarte, S.A. Hughes, S.L. Shapiro, Phys. Rev. D {\bf 60}, 87501,
(1999);
M. Shibata, T.W. Baumgarte,  S.L. Shapiro, Ap. J. {\bf 542}, 453, (2000);  
K. Uryu, M. Shibata, Y. Eriguchi,
Phys. Rev. D {\bf 62}, 104015, (2000);
J.A. Faber, F.A. Rasio, J.B. Manor, Phys. Rev. D {\bf 63}, 044012,
(2001)
\bibitem{ferrarigualtieribor}
V. Ferrari, L. Gualtieri, and A. Borrelli,  Phys. Rev. D {\bf 59}, 124020
(1999)
\bibitem{chandrafer1}
S.Chandrasekhar, V.Ferrari,  Proc. R. Soc. Lond. {\bf A432}, 247
(1990)

\bibitem{zerilli} 
J.F. Zerilli, Phys. Rev D. {\bf 2}, 2141 (1970)

\bibitem{reggewheeler} 
T.Regge, J.A.Wheeler, Phys. Rev. {\bf 108}, 1063
(1957)

\bibitem{nakamurasasaki}
M. Sasaki, T. Nakamura,  Phys. Lett. {\bf 87A}, 85 (1981)

\bibitem{bardeenpress}
J.M.Bardeen, W.H.Press,  J. Math. Phys. {\bf 14}, 7 (1973)

\bibitem{teukolski}
S.A.Teukolsky,  Ap. J. {\bf 185}, 635 (1973)
\bibitem{kojima} 
Y. Kojima,  Prog. Theor. Phys. {\bf 77}, 297, (1987).
\bibitem{ruofflagunapullin}
J. Ruoff, P. Laguna, J. Pullin, Phys. Rev. D {\bf 63}, 064019
(2001)
\bibitem{det}
S.L.Detweiler, Ap. J {\bf 225}, 687 (1978) 
\bibitem{nakaooharakoji}
T.Nakamura, K. Oohara, Y. Kojima,  Prog. Theor.  Phys. Suppl.
{\bf 90}, 1 (1987)
\bibitem{kostas}
V. Ferrari, K.D. Kokkotas,  Phys. Rev. D {\bf 62}, 107504 (2000)
\bibitem{Poisson}
E. Poisson, Phys. Rev. D {\bf 47}, 1497 (1993)
\bibitem{CutlerKennefickPoisson}
C. Cutler, D. Kennefick, E. Poisson, Phys. Rev. D {\bf 50}, 3816 (1994)
\bibitem{BertiFerrari}
E. Berti, V. Ferrari, Phys. Rev. D {\bf 63}, 064031 (2001)
\bibitem{damouriyersathya}
T. Damour, B. R. Iyer, B.S. Sathyaprakash, 
Phys. Rev. D {\bf 57}, 885 (1998)

\end{references}
\end{document}